\documentclass[]{aa}  

\usepackage{graphicx}
\usepackage{txfonts}
\usepackage{lscape}
\usepackage{hyperref}
\hypersetup{colorlinks=true,
            linkcolor=blue,
            urlcolor=blue,
            citecolor=blue}
\usepackage{amsmath}
\usepackage{xcolor}

\begin{document}

   \title{Spectral reconstruction for radiation hydrodynamic simulations of galaxy evolution}

   \titlerunning{Spectral reconstruction for radiation hydrodynamics simulation of galaxy evolution}

   \author{Bernhard Baumschlager\inst{1}
          \and
          Sijing Shen\inst{1}
          \and
          James W. Wadsley\inst{2}
          }
   \authorrunning{B. Baumschlager et al.}

   \institute{Institute of Theoretical Astrophysics, University of Oslo, P.O. Box 1029 Blindern, 0315 Oslo, Norway\\
   \email{bernhard.baumschlager@astro.uio.no}
         \and
             Department of Physics and Astronomy, McMaster University,
             Hamilton, L8S 4M1, Canada\\
             }

   \date{Received ; accepted}

  \abstract{
   Radiation from stars and active galactic nuclei (AGN) plays an important role in galaxy formation and evolution, and profoundly transforms the intergalactic, circumgalactic and interstellar medium (IGM, CGM \& ISM). On-the-fly radiative transfer (RT) has started being incorporated in cosmological simulations, but the complex, evolving radiation spectra are often crudely approximated with a small number of broad bands with piece-wise constant intensity and a fixed photo-ionisation cross-section. 
   Such a treatment is unable to capture the changes to the spectrum as light is absorbed while it propagates through a medium with non-zero opacity.
   This can lead to large errors in photo-ionisation and heating rates.
   In this work, we present a novel approach of discretising the radiation field in narrow bands, located at the edges of the typically used bands, in order to capture the power-law slope of the radiation field. In combination with power-law approximations for the photo-ionisation cross-sections, this model allows us to self-consistently combine radiation from sources with different spectra and accurately follow the ionisation states of primordial and metal species through time. The method is implemented in \textsc{Gasoline2} in connection with \textsc{Trevr2}, a fast reverse RT algorithm with $\mathcal{O}(N_\mathrm{active}\,\log_2\,N)$ scaling. 
    We compare our new piece-wise power-law reconstruction to the piece-wise constant method in calculating the primordial chemistry photo-ionisation and heating rates under an evolving UV-background (UVB) and stellar spectrum, and find that our method reduces errors significantly, up to two orders of magnitude in the case of HeII ionisation.
    We apply our new spectral reconstruction method in RT post-processing of a cosmological zoom-in simulation, MUGS2 g1536 \citep{keller16}, including radiation from stars and a live UVB, and find a significant increase in total neutral hydrogen (HI) mass in the ISM and the CGM due to shielding of the UVB and a low escape fraction of the stellar radiation. This demonstrates the importance of RT and an accurate spectral approximation in simulating the CGM-galaxy ecosystem.
  }


   \maketitle
%
\section{Introduction}

The adoption of on-the-fly RT in hydrodynamic simulations has increased over the past decades, in a wide range of astrophysical applications on scales of galaxies and beyond.
Among those are studies on the impact of photo-electric heating by FUV radiation, photo-ionisation of the interstellar medium by EUV radiation or the escape of ionising radiation in idealised simulations of isolated galaxies \citep[e.g.][]{susa08,kim13,rosdahl15,benincasa20,kannan20,lee20,lee22,katz22a} and zoom-in simulations of galaxy formation and evolution \citep[e.g.][]{pawlik13,kimm17,obreja19,pallottini19,agertz20,pallottini22,garcia23}.
The relative contribution of radiation from stars and AGN on the reionisation of protoclusters has been studied in \citet{treibitsch21}.
The impact of radiation from various sources such as population III stars or high mass x-ray binaries on cosmological structure formation, galaxy evolution and the ISM has been studied by e.g. \citet{wise08,petkova12,wise12,muratov13,jeon14,ricotti16,katz17,rosdahl18,kannan22,rosdahl22,feldmann23,lewis23}.
On the largest scales simulations of cosmological reionisation have been performed by e.g. \citet{iliev06,trac07,trac08,shin08,gnedin14,ocvirk16,rosdahl18,ocvirk20,lewis22}.

\begin{figure*}[t]
    \centering
    \includegraphics[width=\textwidth]{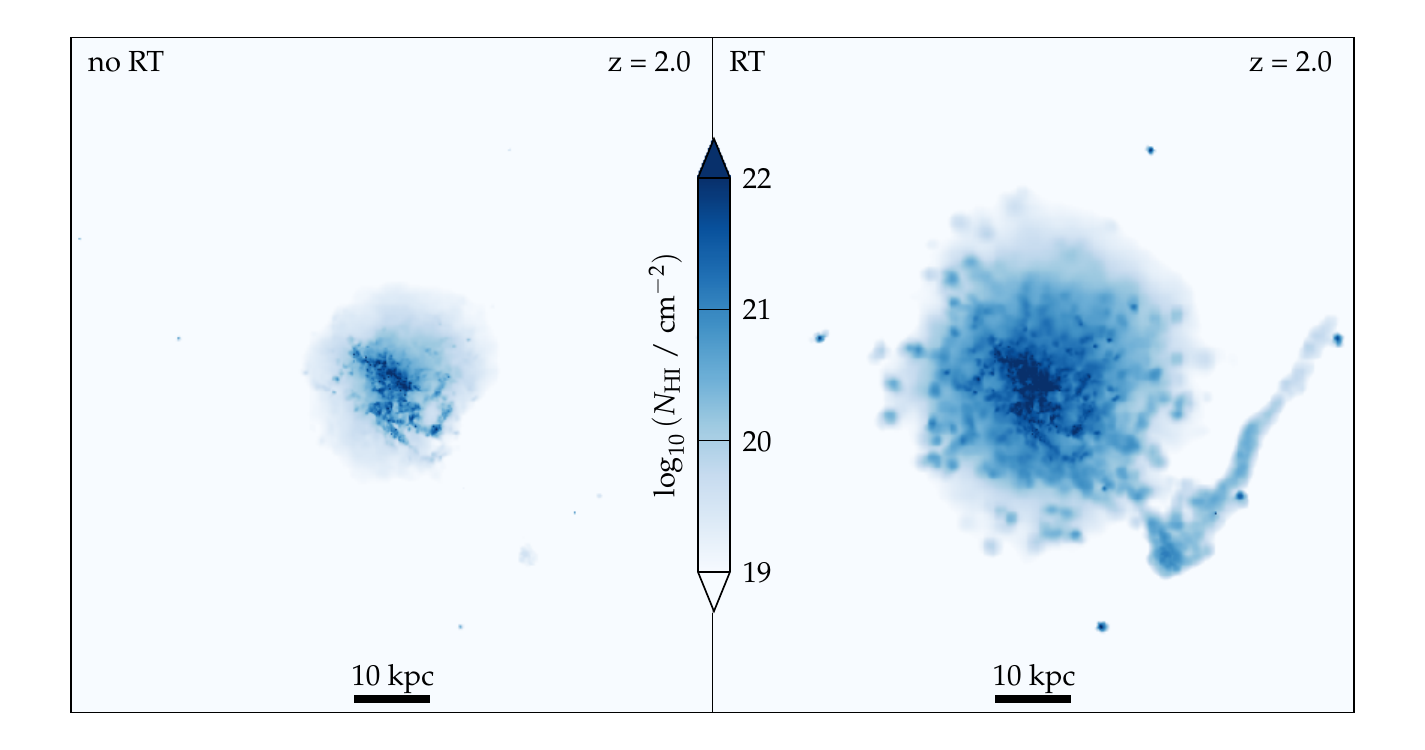}
    \caption{Face-on projection of the neutral hydrogen density ($N_\mathrm{HI}$) in a region with 100 kpc per side centred on the most massive progenitor of g1536 at $z=2$ before (left) and after (right) RT post-processing. See Sect. \ref{sec:mugs} for details.}
    \label{fig:MUGS_HIdens}
\end{figure*}

Due to the complex dependency of the local radiation field on the source spectrum, the spatial distribution of matter and its physical state, as well as the computational expense of radiative transfer calculations, all these simulations rely on simplifications of the radiation field. 
These simplifications are typically (1) the radiation field can be represented by a small number of frequency bands, (2) the radiation within one band is assumed to be monochromatic, i.e. it can be represented by a total intensity at an effective frequency.
A treatment like this is strictly speaking only valid for one specific source spectrum, as the effective monochromatic frequency depends on the shape of the source spectrum.
Additionally, spectral shape changing effects of the intervening matter, e.g. spectral hardening due to stronger absorption at the low energy end of the band, cannot be captured.
Therefore, a treatment like this does not allow for the combination of different types of radiative sources, with different source spectra, like stars of varying ages or combinations of stars with an UVB or AGN, nor is it able to account for the effects of spectral hardening within a radiation band.

Among the aforementioned RT simulations, cosmological zoom-in simulations pose a unique set of challenges, especially when the physical properties of the CGM are studied. 
As the CGM is the interface between IGM and the galaxy itself, it is illuminated from multiple different sources, such as stellar radiation from the central galaxy and its satellite population, as well as distant galaxies and quasars. 
These vastly different sources require the RT algorithm to be able to self-consistently combine their spectra.
In sufficiently large cosmological RT simulation the light emitted by internal sources can be utilised as background radiation from distant sources though the use of periodic boundaries. Due to the high computational cost, it is not practically to achieve a high enough resolution to follow the details of galaxy evolution while simulating a patch of the universe large enough to reproduce a realistic UVB. 
For zoom-in simulations this approach is not possible, as the spectrum would not resemble a representative background spectrum, thus additional sources outside of the zoom region are required to mimic a realistic UVB radiation field.
In order to account for the effect of the UVB on the gas thermo-chemistry, current cosmological simulations without radiative transfer utilise an uniform, instantaneous and redshift dependent treatment of the UVB with \citep[e.g.][]{hopkins14,illustris1,romeel19,agertz21} or without \citep[e.g.][]{eagle,keller16,fiaccion17} corrections for self-shielding. This leads to rapid ionisation of low density gas throughout the simulation volume, once the UVB is switched on, with only high density gas, above the self-shielding threshold or with recombination rates exceeding the UVB photo-ionisation rate, remaining neutral.
Typical cosmological RT zoom-in simulations of galaxy formation, focusing on the ISM of the simulated galaxy, either do not consider ionising radiation from the UVB \citep[e.g.][]{pawlik13,kimm17,garcia23}, or include the effects of UVB in a simplified fashion \citep[e.g.][]{pallottini19,pallottini22, agertz20}, similar to what is used in simulations without RT.  

A treatment akin to those is not sufficient for studies of the CGM and galaxy co-evolution. First, the CGM is highly multiphase with a wide range of densities, ionisation states and metallicities \citep[e.g.][]{tumlinson17, fielding20, decataldo23}. As such, radiation transport, both for stellar/AGN radiation outwards, or background radiation inwards, is likely to have significant spatial variations. Moreover, metals in the CGM are crucial for gas cooling and subsequently impact gas dynamics and galaxy formation. But the ionisation states of metal ions strongly depend on the radiation field, not only the intensity, but also the {\it spectra shape}. For example, one of the major coolants is the fifth ionised oxygen (OVI), which is observed to be abundant in the CGM of star-forming galaxies \citep[e.g.][]{tumlinson13, werk16}. The ionisation potential of OVI is much higher than that of neutral hydrogen, and therefore hard photons from both the UVB and the stellar/AGN radiation are essential \citep[e.g.][]{Oppenheimer18}. Note that the ISM is generally more transparent to higher energy photons and contribution from internal radiation in the CGM closer to galaxies can be significant. While it is common for non-RT (and even RT) simulations to include tabulated metal cooling rates as a function of redshift to capture the evolving, uniform UVB \citep[e.g.][]{wiersma09,shen10, grackle}, such an approach fails to take into account internal radiation from galaxies and the changing spectral shape. Last but not least, observations of the CGM, both in absorption and emission, rely on certain ion tracers (e.g., Ly$\alpha$, CII, CIV, OVI, MgII etc.). An accurate modeling of the radiation field is fundamental for interpret observational data. In this work, we focus on the challenge of modeling the variation of spectra from multiple radiation sources.

We illustrate the effect of RT of ionising photons in zoom-in simulations, including a realistic treatment of the UVB, on the HI distribution in the CGM in Fig. \ref{fig:MUGS_HIdens}, by comparing a zoom-in simulation utilising a spatially uniform UVB (left panel) and after RT post-processing (right panel). The details of the simulation and the RT post-processing are discussed in Sect. \ref{sec:mugs}.
In contrast to previous works, in RT simulations, including a realistic treatment of the UVB, the UVB radiation has to propagate through the volume, whereby it is (partially) absorbed.
The attenuated UVB is not able to penetrate as deep into the CGM, resulting in extended HI disks around galaxies. Additionally, substructures in the CGM, such as ram pressure stripped tails of infalling satellite galaxies or gaseous cores of dwarf satellites, may retain a significant neutral fraction.

These differences highlight the importance of ionising radiation on the evolution of galaxies and the need for full RT simulations, including a realistic treatment of the UVB, in order to fully understand galaxy formation and evolution.

The paper is structured as follows: In Sect. \ref{sec:pwc} we describe commonly used approximations of the radiation field and related factors that affects thermo-chemistry. In Sect. \ref{sec:pwpl} we describe our new approach to reconstruct the spectrum of the incident radiation field during hydrodynamic simulations of galaxy evolution employing on-the-fly RT.
In Sect. \ref{sec:test} we compare our new treatment of the radiation field to commonly used simplifications based on theoretical calculations of photo-ionisation and heating rates and simple test problems and discuss their impact on simulations of galaxies. 
We show results obtained by RT post-processing of a cosmological zoom-in simulation of galaxy evolution with our new spectral reconstruction method in Sect. \ref{sec:mugs}. 
A final discussion is provided in Sect. \ref{sec:discuss}.

\section{Key processes and common simplifications}\label{sec:pwc}

The primary goal of any simulation employing far-UV radiative transfer is to self-consistently account for the interplay of matter with these high energy photons. Among those effects are photo-electric heating of dust, photo-dissociation of molecular hydrogen and photo-ionisation and heating of H and He.

The photo-ionisation rate of an atomic species $i$ can be calculated by:
\begin{equation}\label{eq:ionrate_full}
    \Gamma_{i} = \int{\frac{4\pi\, J_{\rm \nu}}{\mathrm{h}\,\nu}\,\sigma_i\left(\nu\right)\,{\rm d}\nu},
\end{equation}
where $J_{\rm \nu}$ is the radiation field in $\mathrm{ erg\, cm^{-2}\, s^{-1}\, Hz^{-1}\, sr^{-1}}$, $\mathrm{h}$ is Planck's constant and $\sigma_i\left(\nu\right)$ is the frequency dependent ionisation cross-section of species $i$.
The photo-heating rate is given by:
\begin{equation}\label{eq:heating_full}
    \mathcal{H}_{i} = \int{\frac{4\pi\, J_{\rm \nu}}{\mathrm{h}\,\nu}\,\mathrm{h}\left(\nu-\nu_\mathrm{0,i}\right)\,\sigma_i\left(\nu\right)\,{\rm d}\nu},
\end{equation}
where $\mathrm{h}\,\nu_{0,i}$ is the ionisation potential of species $i$.
In the most general case the spectral shape of the incident radiation field $J_\nu$ has a complex dependency on space, time, type of emitting sources and the intervening matter between the emitting source and receiving gas. 
The strong variability of the incident radiation field would require Eqs. \ref{eq:ionrate_full} and \ref{eq:heating_full} to be numerically integrated on a high resolution frequency grid, resulting in prohibitively long calculation times for on-the-fly RT simulations.
In order to solve these equations sufficiently fast and accurately at simulation run-time some simplifications have to be made to the spectral shape, represented by $J_\nu$, and the photo-ionisation cross-section $\sigma(\nu)$.

Early RT simulations of galaxies used a monochromatic approximation for the radiation field. This means all hydrogen ionising photons ($\varepsilon_\gamma\ge13.6\,\mathrm{eV}$) are combined into one band at an effective photon energy \citep[e.g.][]{wise12}.
A more modern approach to simplify the spectral shape is to represent the ionising spectrum by a few, typically three, ionising bands, namely the HI ($13.6-24.6$ eV), HeI ($24.6-54.4$ eV) and HeII ($54.4-\infty$ eV)\footnote{An upper limit of $\infty\,\mathrm{eV}$ for the HeII band is often used to indicate that this band includes all radiation until the highest considered photon energies, while in reality the upper limit is usually around 136 eV, the maximum photon energy considered by stellar population synthesis codes.} 
ionising bands \citep[e.g.][]{pawlik13,rosdahl15,kannan22}.

A second common assumption is the existence of a unique and time independent spectral shape for all different sources of radiation within a simulation. This spectrum is often assumed to be representative of a star forming galaxy which experienced continuous star formation over a few million years \citep[e.g.][]{rosdahl15,pallottini22,kannan22}. 

Considering only the total number of photons in these bands and assuming all sources follow the same spectral shape lets us precalculate effective photo-ionisation cross-sections ($\sigma^{\rm eff}_{i,j}$) for each species $i$ per ionisation band $j$ as:
\begin{equation}\label{eq:sigmaeff}
  \sigma^\mathrm{eff}_{i,j} = \frac{\int \sigma_i(\nu)\,J_\nu\,/\mathrm{h}\nu\, \mathrm{d}\nu}{\int J_\nu\,/\mathrm{h}\nu\, \mathrm{d}\nu},
\end{equation}
where the integration is performed over the energy limits of the considered band. The effective cross-section is then incorporating the shape information of the spectrum.
These assumptions allow Eq. \ref{eq:ionrate_full} to be greatly simplified, for a species $i$, which can be expressed as:
\begin{equation}\label{eq:ionPWC}
  \Gamma_{i} = \sum_{j} N_{j}\, \sigma^\mathrm{eff}_{i,j},
\end{equation}
where $N_j = \int{\frac{4\pi\, J_{\rm \nu}}{\mathrm{h}\,\nu}\,{\rm d}\nu}$ is the photon flux in photons$\,$cm$^{-2}\,$s$^{-1}$ in the ionising band $j$. An analogous expression can be found for the photo-heating rates (Eq. \ref{eq:heating_full}).
We will refer to this discretisation scheme of the radiation filed, as used by the aforementioned RT codes and simulations, as broad band or piece-wise constant (PWC) approximation. 

\begin{figure}[t]
    \centering
    \includegraphics[width=\columnwidth]{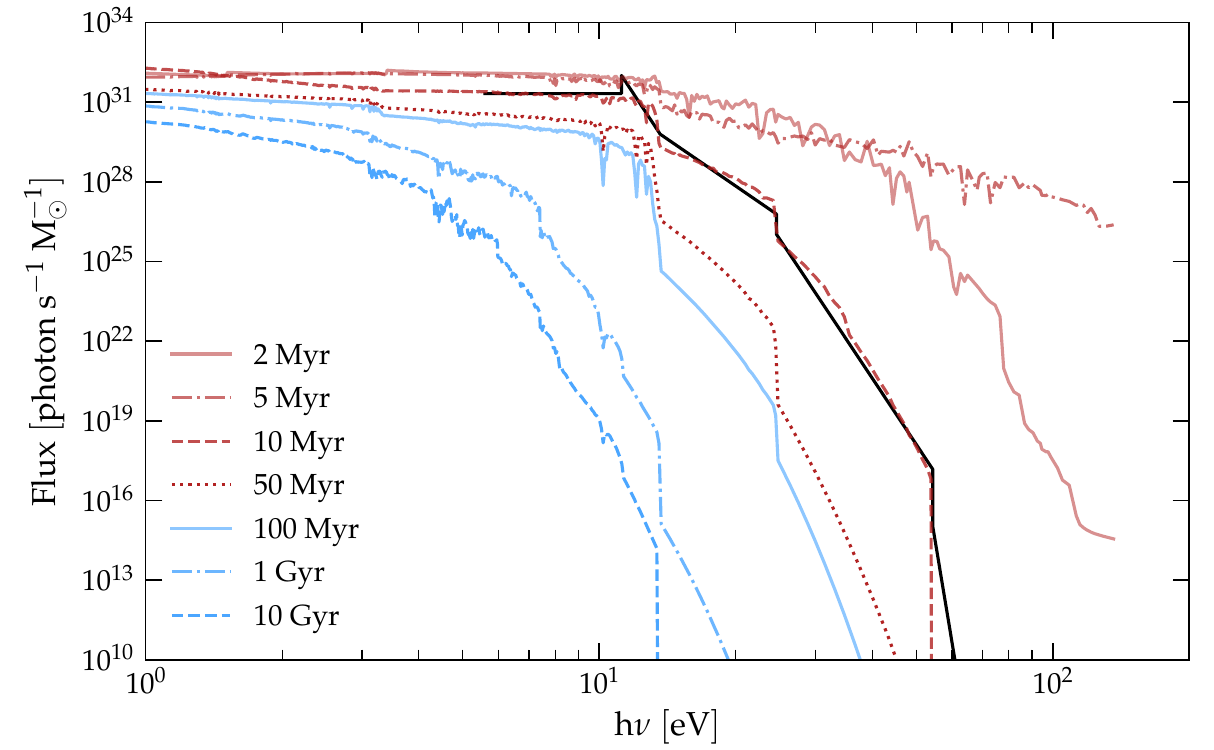}
    \caption{Synthetic stellar population spectra between 1 and 200 eV for different population ages. The spectra are calculated with \textsc{starburst99} \citep{leitherer14} using the Padova AGB evolution tracks with solar metallicity and \citet{kroupa01} IMF. The black line shows the PWPL reconstructed spectrum for a population age of 10 Myr.}
    \label{fig:sb99}
\end{figure}

\begin{figure}[t]
    \centering
    \includegraphics[width=\columnwidth]{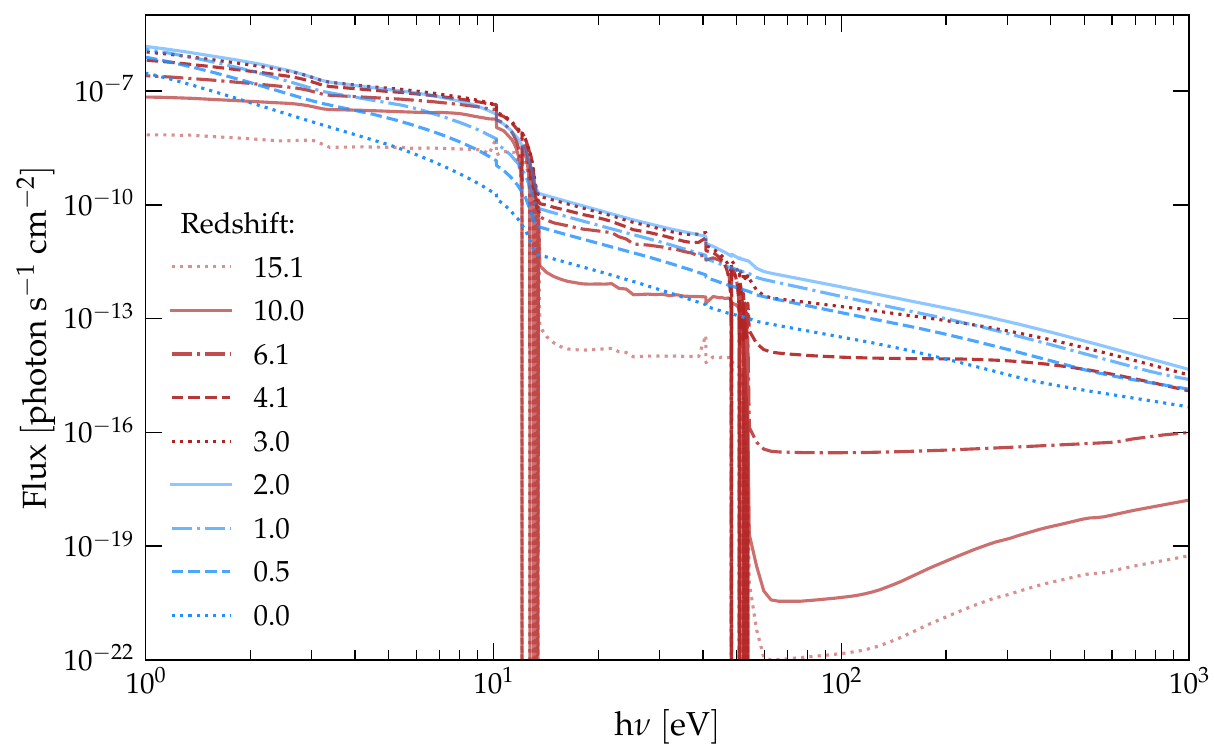}
    \caption{\citet{HM12} UV background spectrum between 1 and 1\,000 eV for various redshifts ranging from $z=15.1$ to $0.0$.}
    \label{fig:hm12}
\end{figure}

The biggest weakness of the PWC approximation is its inability to self consistently combine spectra with different shapes, which is a crucial requirement for high resolution RT simulations of galaxy formation. The local radiation field within a galaxy is highly variable with location, as it depends for example on the age of close by stellar populations and local gas properties. 
Figure \ref{fig:sb99} illustrates the dependency of the stellar source spectrum on the population's age. The synthetic spectra of a simple stellar population (SSP) are calculated with \textsc{starburst99} \citep{leitherer14}, assuming a \citet{kroupa01} initial mass function (IMF), Padova AGB evolution tracks and solar metallicity.
Additional to stellar sources the UVB provides an other important source of ionising photons, in particular for studies of the extended CGM around galaxies or the IGM. Large simulation volumes as used for studies of reionisation 
\citep[e.g.][]{aubert10,shapiro12,gnedin14,lewis22} or cosmological volumes for studies of galaxy evolution \citep[e.g.][]{kannan22} can utilise the light emitted by stars and AGN within the simulation volume to mimic the UVB. 
In cases where only a single galaxy or a small number of galaxies is contained within the simulation volume, as is typical for zoom-in or idealised simulations of galaxies, this approach is not valid. 
Even if the simulation box contains several objects, the resulting background would not reflect a spatially uniform UVB.
Furthermore, the UVB spectrum also changes with time, as illustrated in Fig. \ref{fig:hm12} for the \citet{HM12} UVB.
Both the stellar and UVB source spectra differ by orders of magnitude over time with consequences for key physical processes as shown in Sect. \ref{sec:test}.

Furthermore, spectral hardening due to frequency dependent absorption cannot be fully treated by the PWC approximation.
As radiation is more strongly absorbed at energies close to the ionisation potential of an element, the spectral slope changes. This leads to a mismatch between the spectrum for which $\sigma^\mathrm{eff}_{i,j}$ was precalculated and the actual shape of the local radiation field.
As an example, the hydrogen photo-ionisation cross-sections $\sigma_\mathrm{HI}$ at 13.6 eV and 24.6 eV are 6.35 Mb and 1.24 Mb, respectively, i.e. about a factor of 5 smaller, resulting in a stronger absorption of radiation closer to the low energy edge of the band.
Photo-heating rates are driven by the term $\mathrm{h}\,(\nu-\nu_0)$ in Eq. \ref{eq:heating_full}, that is the excess energy of an ionising photon. This means while ionisations are most effect at energies close to the ionisation energy, due to the larger cross section, the amount of excess energy and thus the heating per ionisation is minimal.
Spectral hardening shifts the effective photon energy in an ionising band to higher energies where it becomes less likely to ionise an atom, but at the same time provides more energy for heating.
An effective frequency/cross-section weighted by photon number generally favours accuracy in ionisation rates, while it fails to reproduce accurate heating rates at the same time.
An alternative and equally valid definition of the effective frequency and cross-section based on a weighting by photon-energy would improve the accuracy of the photo-heating rate, but at the same time degrade it for the ionisation rate.

Additionally, ionisation energies of metals ions, which are used in observation to characterise the physical state of astrophysical gases, are not aligned with the characteristic ionisation energies of H and He.
The PWC approximation is unable to predict the frequency dependent radiation field strength within one band and as such cannot predict the abundance of metal ions, typical tracers of the ISM and CGM, without a priori assumptions on the spectral shape of the incident radiation field.

Some improvements to this approach are already available. For example \textsc{ramses-rt} \citep{rosdahl13} based codes provide the option to recalculate $\sigma^\mathrm{eff}$ and thus the shape information of the spectrum, on-the-fly based on the luminosity averaged spectrum of sources in the simulation volume. This introduces a time dependence of $\sigma^\mathrm{eff}$, but does not consider the spatial distribution of sources.
In order to get a handle on the potentially large errors in the photo-heating rates (see Sect. \ref{sec:instantaneus}) associated with the PWC treatment of the radiation field, these codes used two different effective cross-sections at different effective frequencies. One weighted by photon number (as in Eq. 4) for ionisation and the second weighted by photon energy for heating.

In \textsc{ramses-rtz} \citep{katz22} four ionising UV bands with the band edges at $13.6\,\mathrm{eV}$, $15.2\,\mathrm{eV}$, $24.59\,\mathrm{eV}$, $54.42\,\mathrm{eV}$ and $500\,\mathrm{eV}$ are used. The typical HI ionising band is split into two at 15.2 eV to track H$_2$ photo-ionisation.

Alternatively, \citet{pallottini22} used three narrower bands in the classic HI band ($13.6-24.6$ eV), but neglected ionisation from higher energy photons, and only assumed collisional ionisation for He.
By the approach of splitting the broad bands into several narrower bands it is starting to becomes feasible to follow spectral hardening within the classical broad bands, but it comes at the cost of additional radiation bands. This becomes even more problematic if the broader HeI and HeII ionising bands should be included with a similar spectral resolution as the HI ionising sub-bands.

In a comparison study of numerical methods for computing reionization of intergalactic hydrogen and helium, \citet{leong23} used, depending on the method, between 20 and 200 frequency bins above 13.6 eV in order to ensure an accuracy of a few percent in the frequency integration.

\section{Spectral reconstruction}\label{sec:pwpl}

The aim of our spectral reconstruction method is to simultaneously allow for the combination of different types of radiative sources with varying spectra, be able to calculate precise opacities for the radiation bands, follow spectral hardening within the classic broad ionisation bands, and be analytically integrable so that photo-ionisation and heating rates can be calculated self-consistently and accurately in on-the-fly RT simulations.
In the following section we describe our new discretisation scheme of the photo-ionisation cross-sections and the radiation field, to which we refer to as piece-wise power-law (PWPL) approximation.

\subsection{Photo-ionisation cross-sections}\label{sec:cross}

Fitting functions for the photo-ionisation cross-sections $\sigma$ of atoms are readily available e.g. from \citet{verner96}.
But these fitting functions do not necessarily have an easy to integrate functional form, and would require numerical integration during the calculation of the photo-ionisation and heating rates.

\begin{figure}[t]
    \centering
    \includegraphics[width=\columnwidth]{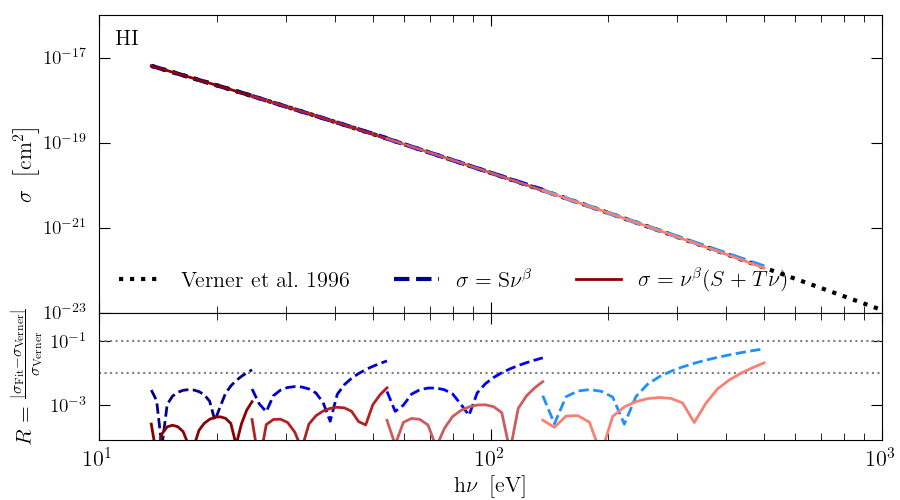}\\
    \includegraphics[width=\columnwidth]{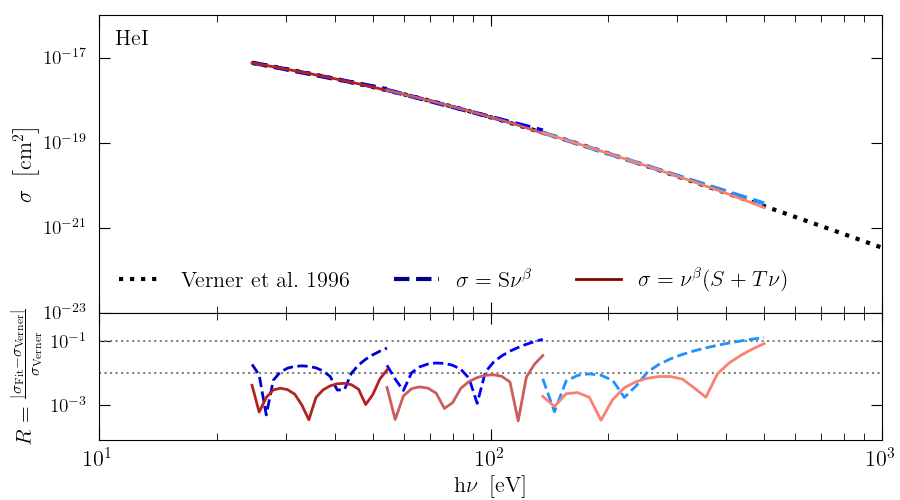}\\
    \includegraphics[width=\columnwidth]{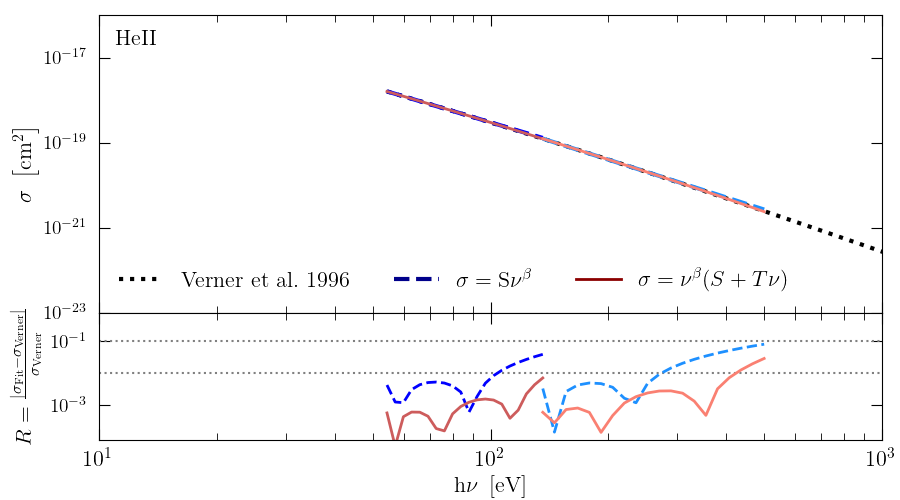}
    \caption{Piece-wise power-law fits to the HI, HeI and HeII photo-ionisation cross-section (top panels) and the associated relative error (lower panels). The black dotted line shows the cross-sections from \citet{verner96}, the blue dashed lines a simple power-law fit and the red solid lines a power law series (Eq. \ref{eq:dpl})
    The horizontal dotted lines in the lower panels indicate relative errors of 1\% and 10\%.}
    \label{fig:cross}
\end{figure}

In order to derive an analytically integrable expression for $\sigma$, we fit truncated power-law series to the fitting functions of \citet{verner96} for each of the broad radiation bands: 
\begin{equation}\label{eq:dpl}
    \sigma_{\rm i,j}(\nu) = \nu^{\beta_{\rm i,j}} \left(S_{\rm{i,j}} + T_{\rm{i,j}}\,\nu\right)~\mathrm{cm}^2,
\end{equation}
where $\beta_{\rm i,j}$, $S_{\rm{i,j}}$ and $T_{\rm{i,j}}$ are the fitting parameters for species $i$ in the broad energy band $j$, and are given in Table \ref{tab:crosssection}.
Figure \ref{fig:cross} shows a comparison of the \citet{verner96} fitting function for the ionisation cross-sections, a piece-wise power law and 
a piece-wise power-law series fit, as well as their associated relative errors for the HI, HeI and HeII photo-ionisation cross-section. As it can be seen form the relative errors, the piece-wise power-law series fit provides significantly lower errors compared to a single power-law. Thereby the relative error generally stays below 1\% and only exceeds this level at the highest energies of some bands, with a maximum error of $\sim10\%$ for $\sigma_\mathrm{He}$ at 500 eV.

\begin{table}
    \centering
    \caption{Piece-wise power-law series fit parameters ($\beta^{\rm i,j}$, $S_{\rm i,j}$ and $T_{\rm i,j}$) for the approximated photo-ionization cross-sections $\sigma$ in cm$^2$ for HI, HeI and HeII ionization in the considered energy bands.}
    \begin{tabular}{l|ccc}
         \multicolumn{4}{l}{HI}\\
         \hline\hline
         Energy band & $\beta_{\rm{i,j}}$ & $S_{\rm{i,j}}$ & $T_{\rm{i,j}}$ \\
         $[\mathrm{eV}]$ &  &  &  \\
         \hline
         13.6 - 24.6   & -2.55 & 5.66$\,\times\,10^{-15}$ & -5.25$\,\times\,10^{-17}$ \\
         24.6 - 54.4   & -2.72 & 8.37$\,\times\,10^{-15}$ & -3.57$\,\times\,10^{-17}$ \\
         54.4 - 136.0  & -2.90 & 1.45$\,\times\,10^{-14}$ & -2.33$\,\times\,10^{-17}$ \\
         136.0 - 500.0 & -3.07 & 2.88$\,\times\,10^{-14}$ & -1.37$\,\times\,10^{-17}$ \\
         \hline
         \multicolumn{4}{l}{ }\\
         \multicolumn{4}{l}{HeI}\\
         \hline\hline
         13.6 - 24.6   & 0.00  & 0.00 & 0.00  \\
         24.6 - 54.4   & -1.36 & 7.46$\,\times\,10^{-16}$ & -6.65$\,\times\,10^{-18}$ \\
         54.4 - 136.0  & -1.93 & 4.87$\,\times\,10^{-15}$ & -2.01$\,\times\,10^{-17}$ \\
         136.0 - 500.0 & -2.73 & 1.30$\,\times\,10^{-13}$ & -1.23$\,\times\,10^{-16}$ \\
         \hline
         \multicolumn{4}{l}{ }\\
         \multicolumn{4}{l}{HeII}\\
         \hline\hline
         13.6 - 24.6   & 0.00  & 0.00 & 0.00  \\
         24.6 - 54.4   & 0.00  & 0.00 & 0.00  \\
         54.4 - 136.0  & -2.58 & 5.34$\,\times\,10^{-14}$ & -1.08$\,\times\,10^{-16}$ \\
         136.0 - 500.0 & -2.82 & 1.38$\,\times\,10^{-13}$ & -8.55$\,\times\,10^{-17}$ \\
         \hline
    \end{tabular}
    \label{tab:crosssection}
\end{table}

\subsection{Discretising the source spectra}\label{sec:raddisc}

Observing typical source spectra as shown in Figures \ref{fig:sb99} and \ref{fig:hm12}, two main common features of their shapes becomes obvious. The first is the presence of jumps in intensity across the ionisation edges of H and He, and second is the relatively constant slopes between these ionisation edges at a given time.

In order to capture potential jumps across the ionisation edges we use two narrow ($\Delta \varepsilon_\gamma=0.1\,\mathrm{eV}$) energy bins, just to the left and right of the ionisation energies of hydrogen and helium. Between these narrow bands we assume that the spectral shape can be described as power-law. The slope of the individual power-law pieces is calculated on-the-fly for each receiving resolution element. In practise, we first add the attenuated photon fluxes within the narrow bands, and calculate the spectral slope based on the combined intensities.
This approach naturally follows spectral hardening within the typical ionisation bands, as the fluxes within the narrow bands are attenuated according to their respective opacity. In contrast, in the PWC approximation the intensity within one ionising band is calculated based on a single effective opacity and thus is not able to capture any changes in the shape of the radiation filed due to frequency dependent absorption.

This leads to 6 photon bands in the range of 13.6 to 136 eV.
We chose 136 eV as upper energy limit for the HeII band (HeII high), as this is the typical upper energy limit of popular stellar spectral synthesis codes, such as \textsc{starburst99} \citep{leitherer14} or \textsc{slug} \citep{dasilva12}. 
An upper energy limit of 136 eV for the source spectra is insufficient to accurately capture the photo-ionisation and heating rates of helium, in particular HeII. This is especially true in cases were the high-energy spectral slope is positive, such as for the \citet{HM12} UVB at $z \gtrsim 4$, or in cases of strong spectral hardening.
In order to improve the accuracy of the calculated HeII photo-ionisation rates, we include an additional high energy (XUV) band at 500 eV.
We additionally include one FUV band ($5.6-11.2$ eV) and two Lyman-Werner bands for photo-electric heating and for future inclusion of H$_2$ chemistry, resulting in a total of 10 energy bands. A comprehensive list of the used bands is given in Table \ref{tab:bands}.

\begin{table}
    \centering
    \caption{Radiation bands used in the PWPL approximation. The columns list the upper and lower energy bounds of the bands and their representative photon energy.}
    \begin{tabular}{lccc}
        \hline\hline
         Band & $\varepsilon_{\rm \gamma,\,min}$ & $\varepsilon_{\rm \gamma,\,max}$ & $<\varepsilon_{\rm \gamma}>$ \\
         & [eV] & [eV] & [eV] \\
         \hline
         FUV       & 5.6  & 11.2 &  8.4  \\
         LW low    & 11.2 & 11.3 & 11.25 \\
         LW high   & 13.5 & 13.6 & 13.55 \\
         HI low    & 13.6 & 13.7 & 13.65 \\
         HI high   & 24.5 & 24.6 & 24.55 \\
         HeI low   & 24.6 & 24.7 & 24.65 \\
         HeI high  & 54.3 & 54.4 & 54.35 \\
         HeII low  & 54.4 & 54.5 & 54.45 \\
         HeII high & 135.9 & 136.0 & 135.95 \\
         XUV       & 499.9 & 500.0 & 499.95 \\
         \hline
    \end{tabular}
    \label{tab:bands}
\end{table}

\subsection{Calculating photo-ionization \& heating rates}\label{sec:rates}

Combining our narrow band power-law approximations for the radiation field and ionization cross-sections allows us to simplify Eqs. \ref{eq:ionrate_full} and \ref{eq:heating_full} into:
\begin{equation}\label{eq:ionrateApStep}
    \Gamma_{i} = \int{F_{j}\,\nu^{\alpha_j} \,\left( S_{i,j}\,\nu^{\beta_{i,j}} + T_{i,j}\,\nu^{\beta_{i,j}+1}\right) \,d\nu} ,
\end{equation}
and
\begin{equation}\label{eq:heatrateApStep}
    \mathcal{H}_{i} = \int{F_j\, \nu^{\alpha_j}\, \mathrm{h}\left(\nu-\nu_{0,i}\right)\,\left[ S_{i,j}\,\nu^{\beta_{i,j}} + T_{i,j}\,\nu^{\beta_{i,j}+1}\right]\,d\nu},
\end{equation}
here $F_j$ is the photon flux in ${\rm photons\, cm^{-2}\, s^{-1}\, Hz^{-1}}$ and $\alpha_j$ is the slope of the power-law spectrum within the broad band j, $S_{i,j}$, $T_{i,j}$ and $\beta_{i,j}$ are the fitting parameters of the power-law photo-ionisation cross-sections, given in Table \ref{tab:crosssection}, and $\nu_{0,i}$ is the ionisation potential of species i.
These expression can be evaluated analytically for each broad band, leading to a summation over these bands. Therefore Eqs. \ref{eq:ionrateApStep} and \ref{eq:heatrateApStep} become:
\begin{equation}
\begin{aligned}\label{eq:ionrateAp}
    &\Gamma_{i} = \sum_{j} F_j \,\times \\
    &\left\{ S_{i,j} \left[\frac{\nu_{{\rm max},j}^{\alpha_j+\beta_{\rm i,j}+1} - \nu_{{\rm min},j}^{\alpha_j+\beta_{i,j}+1}}{\alpha_j+\beta_{\rm i,j}+1}\right] + T_{i,j} \left[\frac{\nu_{{\rm max},j}^{\alpha_j+\beta_{\rm i,j}+2} - \nu_{{\rm min},j}^{\alpha_j+\beta_{\rm i,j}+2}}{\alpha_j+\beta_{\rm i,j}+2}\right] \right\}
\end{aligned}
\end{equation}
and
\begin{equation}
\begin{aligned}\label{eq:heatrateAp}
    \mathcal{H}_{i} = \sum_{j}\, F_j\,\mathrm{h}\,\times\,& \Biggl\{ T_{i,j}\left[ \frac{\nu_{{\rm max},j}^{\alpha_j+\beta_{i,j}+3}- \nu_{{\rm min},j}^{\alpha_j+\beta_{i,j}+3}}{\alpha_j+\beta_{i,j}+3} \right] \\ & + \left(S_{i,j} - T_{i,j}\nu_{0,i} \right) \left[\frac{\nu_{{\rm max},j}^{\alpha_j+\beta_{i,j}+2}- \nu_{{\rm min},j}^{\alpha_j+\beta_{i,j}+2}}{\alpha_j+\beta_{i,j}+2} \right] \\ & - S_{\rm i,j}\,\nu_{0,i}\, \left[ \frac{\nu_{{\rm max},j}^{\alpha_j+\beta_{i,j}+1} - \nu_{{\rm min},j}^{\alpha_j+\beta_{i,j}+1}}{\alpha_j+\beta_{i,j}+1} \right] \Biggl\} ,
\end{aligned}
\end{equation}
where $\nu_{{\rm min},j}$ and $\nu_{{\rm max},j}$ are the lower and upper frequency bound of ionising band $j$ as indicated in column one of Table \ref{tab:crosssection}.
The approximations \ref{eq:ionrateAp} and \ref{eq:heatrateAp} are considerably faster to evaluate than Eqs. \ref{eq:ionrate_full} and \ref{eq:heating_full} as the integral does not need to be numerically evaluated on a high frequency resolution grid.

Unlike to the PWC approximation, with the PWPL spectral reconstruction it is possible to calculate the intensity of the radiation field at any frequency between the band edges.
This enables us to calculate precise ionisation rates of for metal lines.

\section{Model validation}\label{sec:test}

\subsection{Instantaneous ionisation and heating rates}\label{sec:instantaneus}

In order to validate the PWPL approximation we compute the instantaneous photo-ionisation and heating rates for HI, HeI and HeII according to Eqs. \ref{eq:ionrateAp} and \ref{eq:heatrateAp} and compare the results to the numerical integral of Eqs. \ref{eq:ionrate_full} and \ref{eq:heating_full} in the energy range $13.6\leq \mathrm{h}\nu/\mathrm{eV}\leq500$ and to the PWC approximation (Eqs. \ref{eq:ionrateApStep} and \ref{eq:heatrateApStep}).
We perform this test for two typical source types with time depended spectra, namely the UVB and a $10^6\,\mathrm{M}_\odot$ star cluster at a distance of 100 pc.
\begin{figure*}[t]
    \centering
    \includegraphics[width=0.48\textwidth]{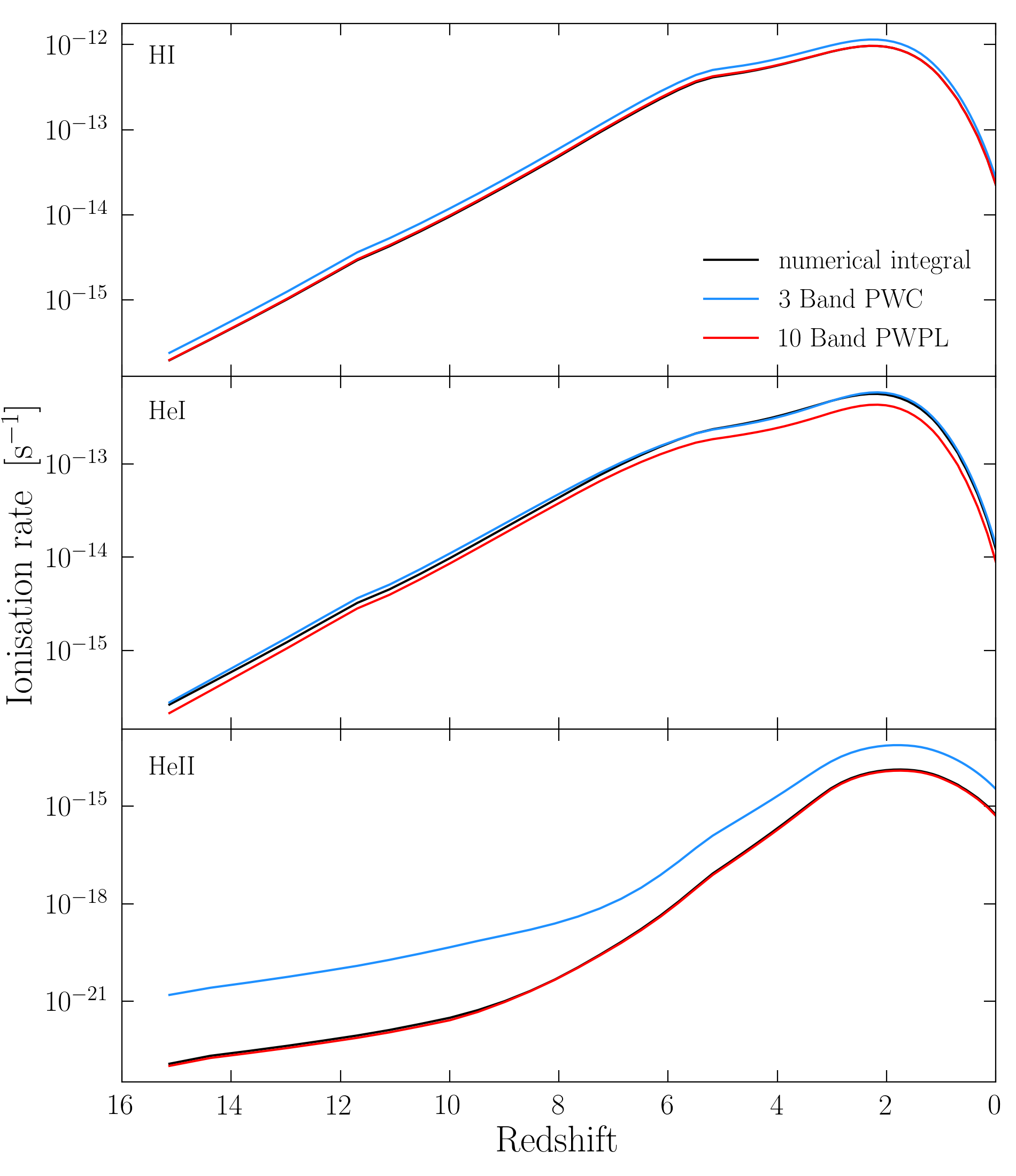}
    \hfill
    \includegraphics[width=0.48\textwidth]{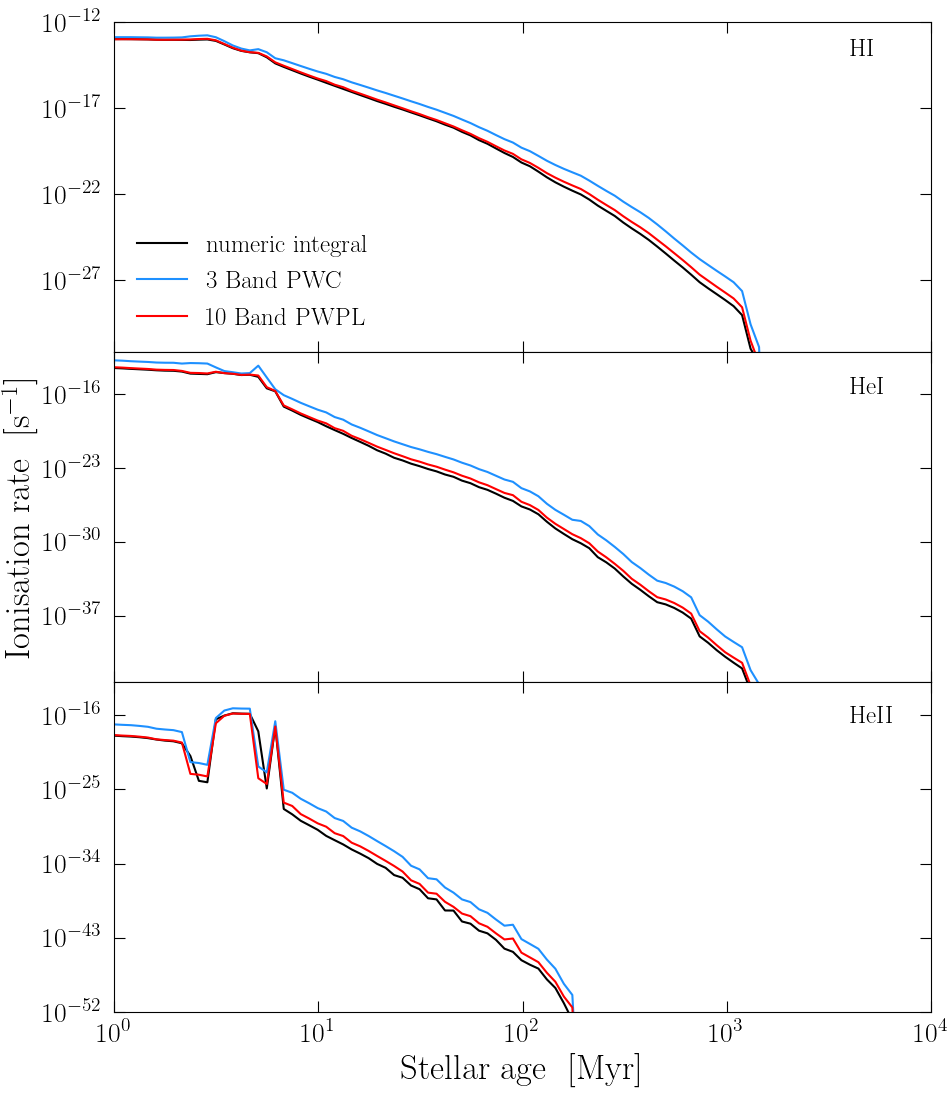}
    \\
    \includegraphics[width=0.48\textwidth]{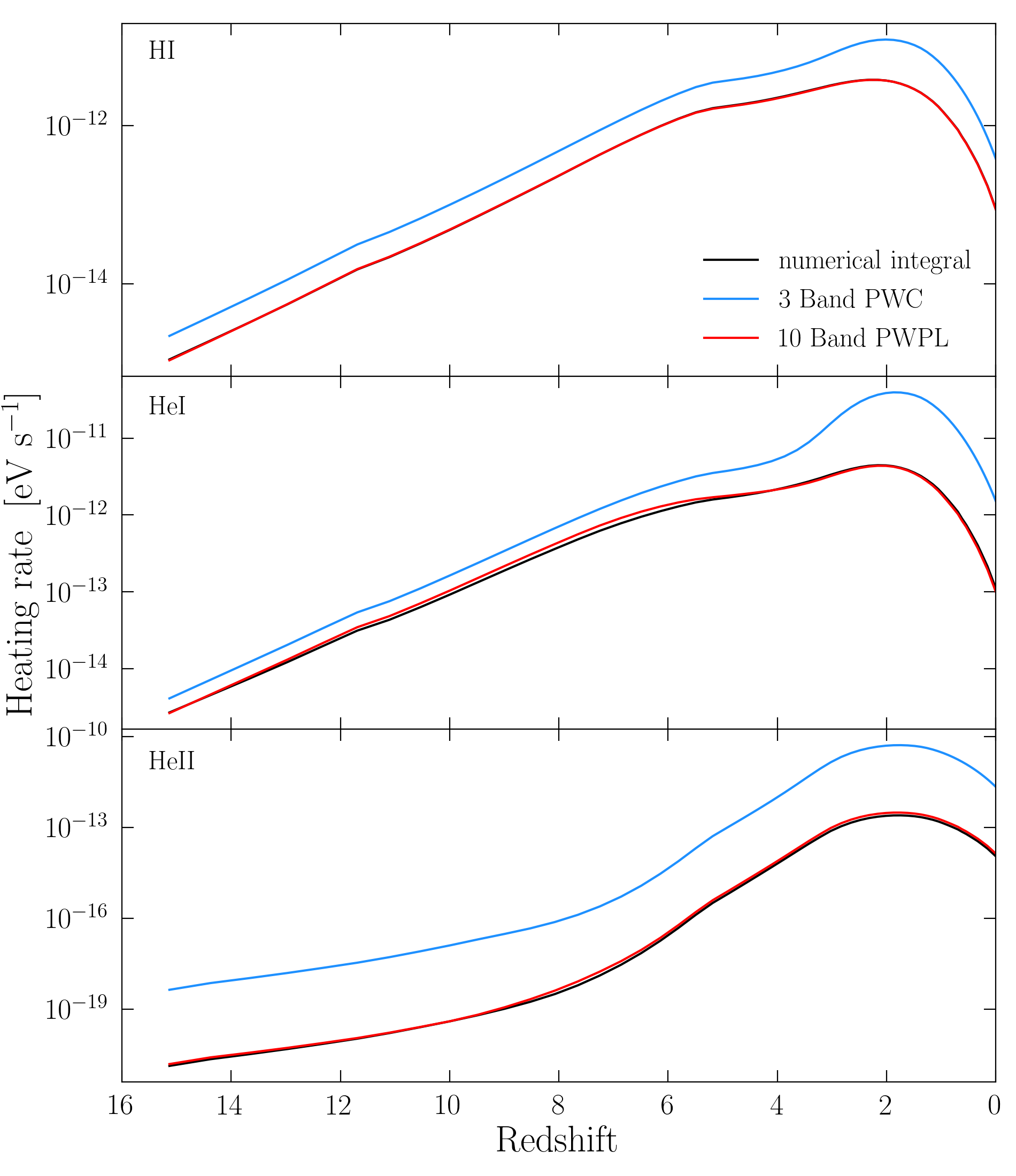}
    \hfill
    \includegraphics[width=0.48\textwidth]{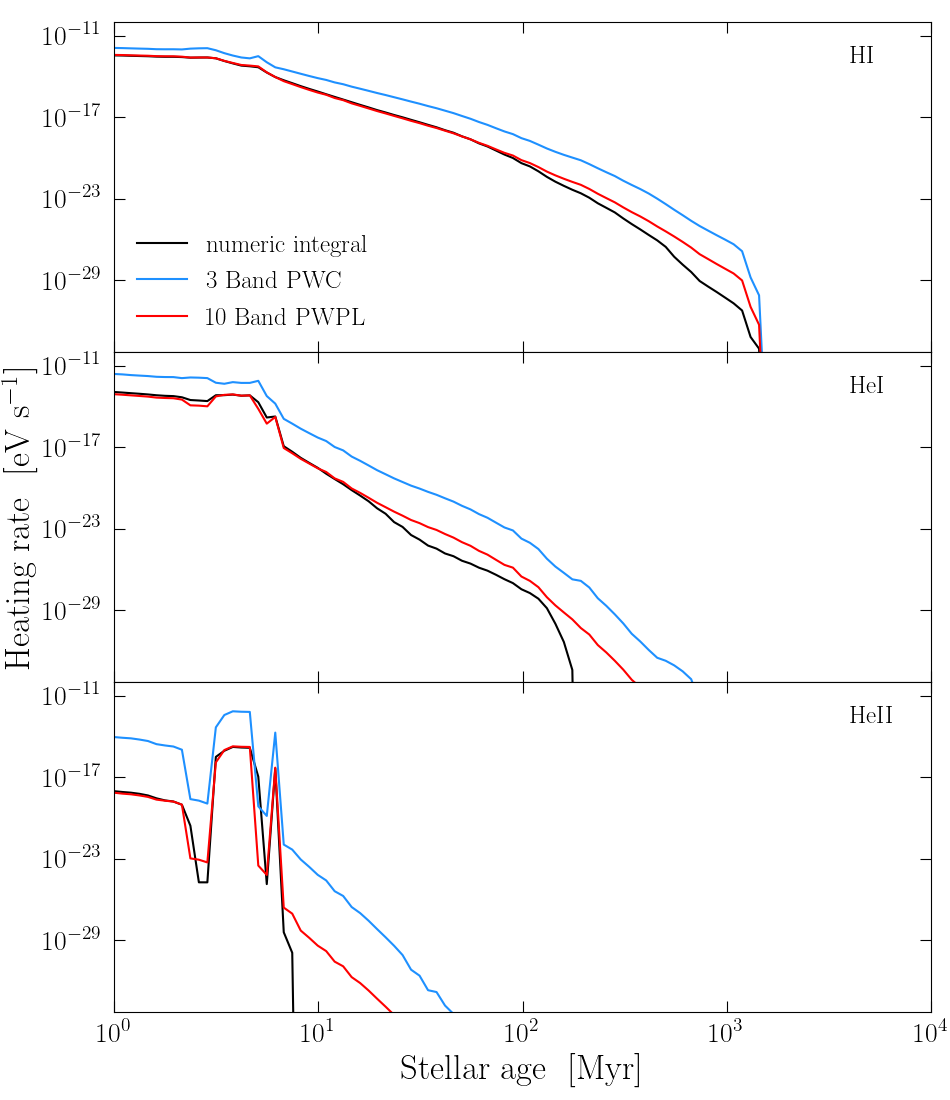}
    \caption{Photo-ionisation rate (top) and photo-heating rate (bottom), for a redshift dependent uniform \citet{HM12} UVB spectrum (left) and an aging SSP with $M_\mathrm{cl}=10^6\,\mathrm{M}_\odot$ at a distance of 100 pc (right). The blue line shows the ionisation and heating rate for the common PWC approximation of the radiation field in combination with a constant $\sigma_\mathrm{eff}$. The red line corresponds to our new PWPL approximation for the radiation field in combination with the power-law photo-ionisation cross-section (Eq. \ref{eq:ionrateApStep}).}
    \label{fig:validation}
\end{figure*}
For the first case we use the \citet{HM12} UVB spectrum over the full published redshift range from $z=15~ \mathrm{to}~ 0$.
For the second case we calculate the intensities of the stellar sources in each bin based on synthetic spectra of a SSP derived with \textsc{starburst99} \citep{leitherer14} using the Padova AGB stellar evolution tracks for solar metallicity. We assume the population follows the \citet{kroupa01} IMF with an lower and upper mass limit of 0.1 and 100 $\mathrm{M}_\odot$, respectively, and ages between 1 Myr and 13 Gyr. 
As the original source spectra are not actual power laws, the integrated number of photons between two narrow bands in the PWPL approximation is not necessarily equal to the total number of the corresponding broad band. To ensure conservation of photons between the broad band and the power-law approximation, we scale the intensities in the narrow bands, while keeping the power-law slope constant.
The black line in Fig. \ref{fig:sb99} shows the reconstructed spectrum of the stellar population at an age of 10 Myr.

For the PWC calculations, we use 3 ionising bands with their boundaries at 13.6, 24.6, 54.4 and 500 eV and the same effective photo-ionisation cross-sections weighted by photon number as used in \citet{kannan22}. 

We compare the photo-ionisation (top panels) and heating (bottom panel) rates for the \citet{HM12} UVB (left) and the SSP (right) calculated with the PWPL (red lines), PWC (blue lines) to the numerical integral of Eqs. \ref{eq:ionrate_full} and \ref{eq:heating_full} (black lines) in Fig. \ref{fig:validation}.
For both test cases the PWPL follows the shape of the numerical integral closely, while the PWC approximation has the tendency to produce higher ionisation and heating rates.

\begin{figure}[t]
    \centering
    \includegraphics[width=\columnwidth]{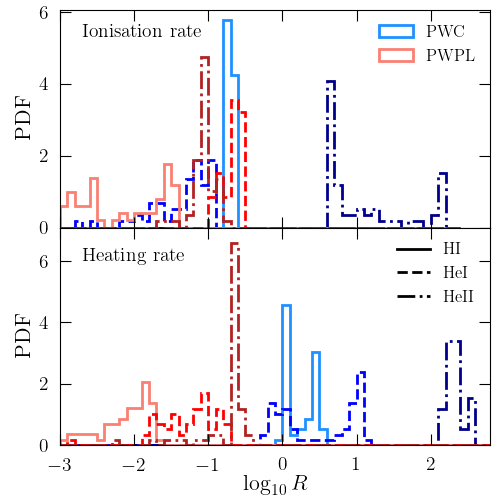}
    \caption{Relative error distribution for the ionisation (top) and heating (bottom) rates for the redshift dependent UVB spectrum, for the PWPL (reds) and PWC (blues) approximation. Different line styles show the error distribution for HI (solid), HeI (dashed) and HeII (dot-dashed). The errors are calculated relative to the numerical integration of Eqs. \ref{eq:ionrate_full} and \ref{eq:heating_full}.}
    \label{fig:errorUVB}
\end{figure}

The PWPL approximation simultaneously predicts the photo-ionisation and heating rates accurately in the UVB case, with mean errors of 0.005, 0.11 and 0.04\,dex for HI, HeI and HeII ionisation rates, respectively, and 0.004, 0.03 and 0.08\,dex for the heating rates.
As apparent from the left panels of Fig. \ref{fig:validation}, the PWC approximation fails to predict photo heating rates in this test, with mean errors of 0.44, 0.66 and 2.32\,dex for HI, HeI and HeII photo heating rates, respectively. The mean erros for the photo-ionisation rates of the three species are 0.08, 0.02 and 1.15\,dex.
We show the distribution of relative errors for the PWC (blue lines) and PWPL (red lines) approximations relative to the numerical integral of Eqs. \ref{eq:ionrate_full} and \ref{eq:heating_full}, for the ionisation (top panel) and heating (bottom panel) rates of HI (solid), HeI (dashed) and HeII (dot-dashed) for the redshift dependent UVB case in Fig. \ref{fig:errorUVB}.
We define the relative error as:
\begin{equation}\label{eq:relErr}
    R = \frac{\left|X_\mathrm{approx}-X_\mathrm{true}\right|}{X_\mathrm{true}},
\end{equation}
where X is either the photo-ionisation or heating rate, we assume the numerically integrated values represent the true rates.

As the SSP ages, its spectral shape changes, which cannot be captured by the precalculated cross-section in the PWC case. This results in a varying accuracy, where the deviation from the expected rates increases with population age, for the chosen precalculated cross-sections.
The calculated photo-ionisation and heating rates of the PWPL approximation closely follow the expected values (Fig. \ref{fig:validation} right). Whereas the PWC approximation overestimates both by several orders of magnitude.
The spectral shape changes during the evolution of the SSP cannot be captured by the precalculated cross-sections in the PWC case. 
This results in a population age dependent accuracy, where the deviation from the expected rates increases with population age, for the chosen precalculated cross-sections.
As the population ages, the number of emitted high energy photon decreases, starting at the highest energies and successively progressing to lower energies (see Fig. \ref{fig:sb99}). 
Due to the discretisation of the source spectrum, for both the PWC and PWPL methods, the highest energy bands still hold photons, including at energies where the original sources emit little to no photons. 
As the heating rate is governed by the excess energy per ionisations, i.e. the term $\mathrm{h}\left(\nu-\nu_\mathrm{0,i}\right)$ in Eq. \ref{eq:heating_full}, even a small number of high energy photons can significantly increase the photo-heating rate.
A behaviour like this has to be expected for any approximation of the radiation field, that relies on a limited number of radiation bands.
This causes huge nominal errors, but the effect is negligible as at this point the heating rates are already $\sim10$ orders of magnitude lower compared to a young stellar population. This effect is particularly apparent in the HeII ionisation rates for population ages $\gtrsim8\,\mathrm{Myr}$ (bottom right panel of Fig. \ref{fig:validation}).

We note that the behaviour of the PWC approximation could be improved by carefully evaluating $\sigma_{i,j}^\mathrm{eff}$ for each of the different spectral shapes in the test. 
This approach is not practical in full scale RT simulations, even in this limited test with combined 160 different spectra the time needed to calculate all 960 cross-sections by far exceeds the time needed to calculate the ionisation and heating rates.

\subsection{Expanding HII-region}

The expansion of an HII-region in pure hydrogen is a typical test case used to asses the performance of radiative transfer codes \citep[e.g.][]{gnedin01,iliev06a,iliev09,rosdahl13,grond19,kannan19}. 
Thereby, the density is assumed to be constant, the radiation field is represented by a single ionising band and recombination is limited to case B. In the classic Str\"omgren sphere test the temperature is kept constant, although usually a second test with varying temperature is performed as well.

We implemented our new PWPL spectral reconstruction method for the radiation field into an early test version of \textsc{Trevr2} \citep{wadsley23}, the new radiative transfer method of the Tree-SPH code \textsc{Gasoline2} \citep{wadsley17}. \textsc{Trevr2} is a fast reverse ray tracing algorithm with $\mathcal{O}(N_\mathrm{active}\, \log_2\, N_\mathrm{src})$ computational cost, here $N_\mathrm{active}$ is the number of currently active receiving particles and $N_\mathrm{src}$ is the number of sources. It performs the ray tracing from the receiver to the sources along HEALPix \citep{gorski05} cones.

We perform a multi-band version of the expanding HII-region test with variable temperature, dropping the restriction of only having one radiation band and do not restrict recombination to case B. Additional to the pure hydrogen case, we also present results for a primordial gas mixture of hydrogen and helium, with a helium mass fraction of 23.6\%.
A single source of photons is placed in the centre of the simulation box with a side length of $16\,\mathrm{kpc}$. We use $64^3$ gas particles to model the constant density medium with $n_\mathrm{H}=10^{-3}\,\mathrm{cm}^{-3}$ and an initial gas temperature of $T=100\,\mathrm{K}$. For the primordial mixture case we keep $n_\mathrm{H}=10^{-3}\,\mathrm{cm}^{-3}$ and add mass to the particles so that the helium mass fractions reaches the aforementioned 23.6\%, resulting in $n_\mathrm{H+He}\approx 1.07\times 10^{-3}\,\mathrm{cm}^{-3}$.
Following \citet{katz22}, we use a blackbody with $T_\mathrm{BB}=10^{4.759}\,\mathrm{K}$ and a luminosity of $L_\mathrm{BB}=10^{4.324}\,\mathrm{L_\odot}$ as source, which resembles a $15\,\mathrm{M_\odot}$ main-sequence star \citep{schaerer02}.
As for the previous tests, we scaled the number of photons in the narrow PWPL bands so that the total number of photons in the broad bands equals the integrated number of photons in the same energy interval within the PWPL approximation, while keeping the power law slope constant.

We compare our PWPL results to the PWC approximation and comparison runs from \textsc{cloudy} \citep[version 17.02,][]{ferland17}. 
For the PWC runs we calculate $\sigma_{i,j}^\mathrm{eff}$ according to Eq. \ref{eq:sigmaeff} for the considered blackbody spectrum.
For the \textsc{cloudy} calculations we deactivate molecular chemistry and dust physics as both are not considered with in the chemical network of \textsc{Gasoline2}.

\begin{figure*}[t]
    \centering
    \includegraphics[width=\textwidth]{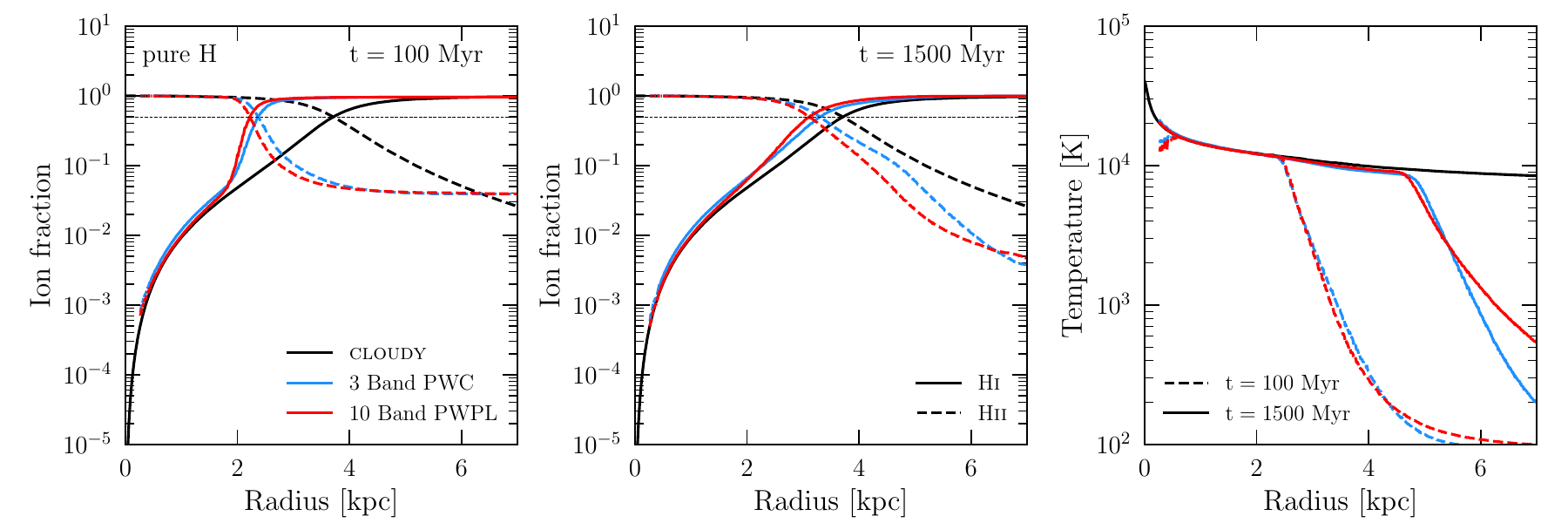}
    \caption{Expanding HII region in a pure hydrogen environment with $n_\mathrm{H}=10^{-3}\, \mathrm{cm}^{-3}$ and an initial temperature of $100\,\mathrm{K}$. The hydrogen ion fractions (HI solid and HII dashed lines) after $t=100~(1500)~\mathrm{Myr}$ is shown in the left (middle) panel. Red lines show the results using the 10 Band PWPL, blue lines using the 3 Band PWC approximation and the black lines represent the \textsc{cloudy} equilibrium solution. The thin dashed line represents an ion fraction of 50\%. The right panel shows the temperature profile of the HII region after $t=100$ and $1500~\mathrm{Myr}$ as dashed and solid lines, respectively.
    }
    \label{fig:Honly}
\end{figure*}

\begin{figure*}[t]
    \centering
    \includegraphics[width=\textwidth]{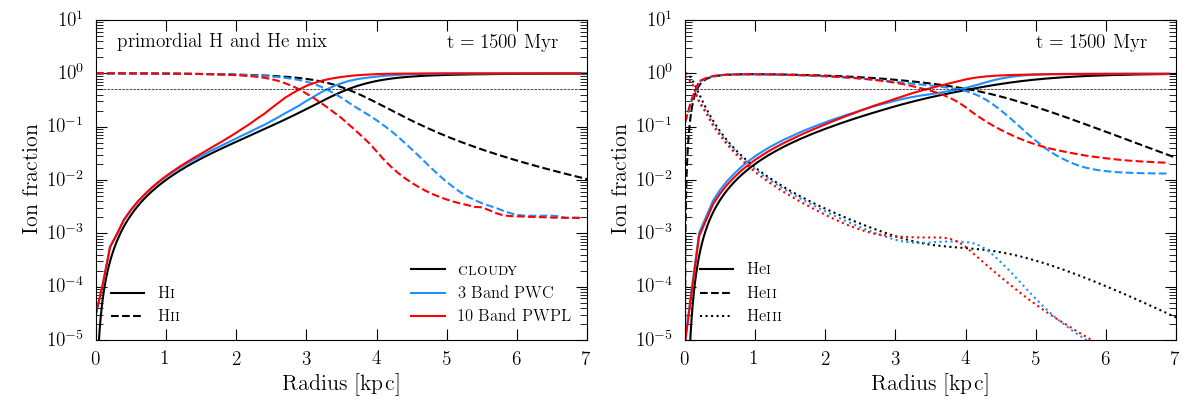}
    \caption{Expanding HII region in a primodial mixture of hydrogen and helium environment with $n_\mathrm{H}=10^{-3}\, \mathrm{cm}^{-3}$ and an initial temperature of $100\,\mathrm{K}$. The hydrogen ion fractions (HI solid and HII dashed lines) after $t=1500~\mathrm{Myr}$ is shown in the left panel. The right panel shows the helium ion fractions (HeI solid, HeII dashed and HeIII dotted lines) after $t=1500~\mathrm{Myr}$. The thin dashed line represents an ion fraction of 50\%.
    } 
    \label{fig:HHe}
\end{figure*}

In Fig. \ref{fig:Honly} we show the resulting hydrogen ion fractions during the rapid expansion phase of the HII region, after $t=100\,\mathrm{Myr}$ (left) and after $t=1500\,\mathrm{Myr}$ when the final equilibrium has been reached. 
The right panel shows the temperature profile of the HII region at the same times. It closely follows the \textsc{cloudy} comparison run until the knee of the temperature profile, where the photo-heated region transitions to the initially cold surrounding medium. 
This difference in the temperature profile at large distances beyond approximately 5 kpc from the central source is a result of inherently different approaches to the chemistry solver. \textsc{Gasoline2} uses a non-equilibrium chemical network to heat the initially cold (100 K) gas, while \textsc{cloudy} directly calculates the equilibrium state.
In Fig. \ref{fig:HHe} we show equilibrium hydrogen (left) and helium (right) ion fraction.

\begin{figure*}[t]
    \centering
    \includegraphics[width=\textwidth]{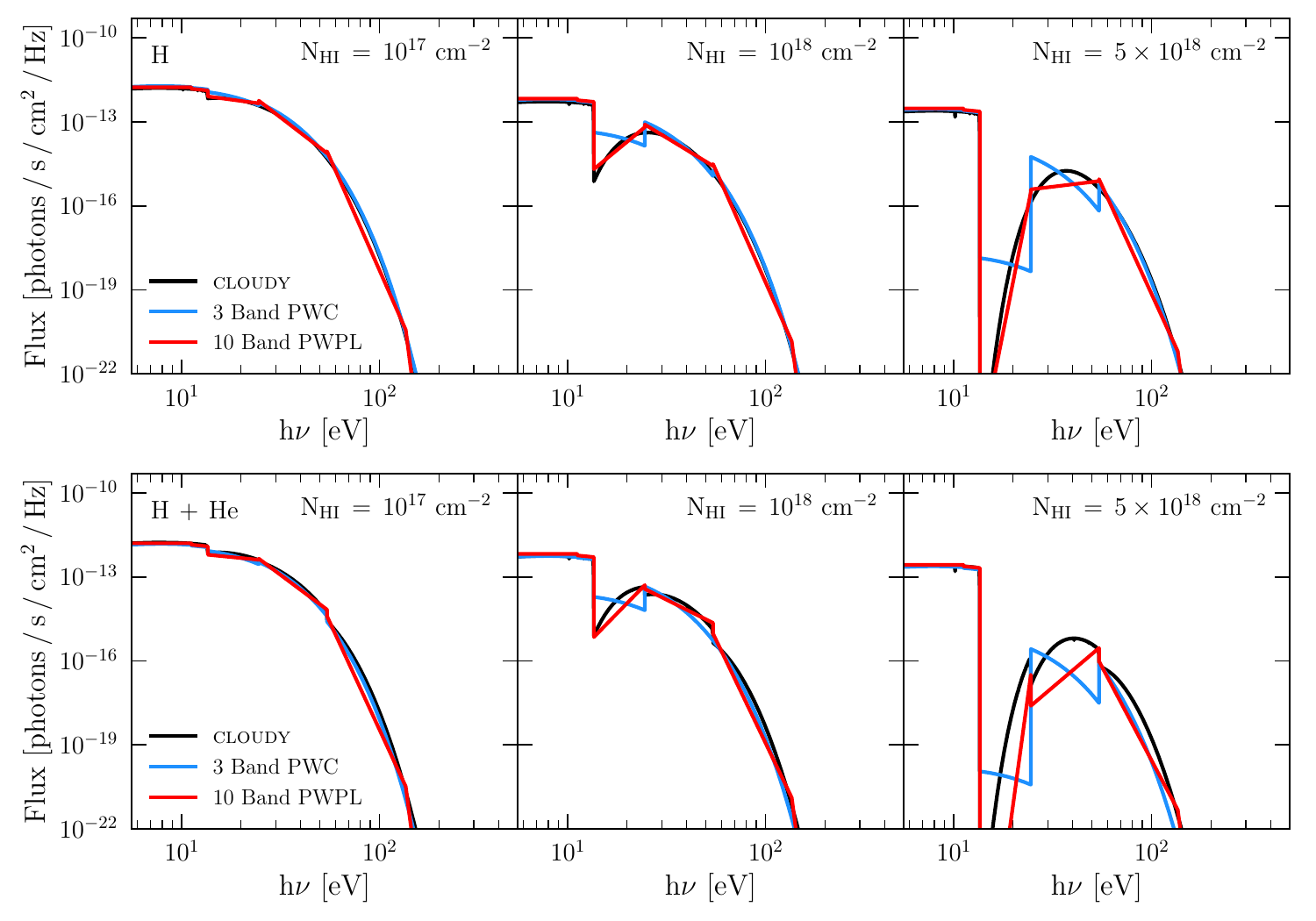}
    \caption{Reconstructed radiation field after $t=1500~\mathrm{Myr}$, at column densities of $N_\mathrm{HI}=10^{17},~10^{18}~\mathrm{and}~5\times10^{18}~\mathrm{cm}^{-2}$ (from left to right) from the source, for the pure hydrogen (top) and primordial hydrogen and helium mixture (bottom) case. The black line denotes the \textsc{cloudy} reference runs, the red lines show the reconstructed spectrum using the PWPL approximation and the blue line represents the classic 3 band PWC approach. For illustrative purposes, we choose to represent the spectral shape of the 3 band PWC case as scaled intervals of the initial blackbody spectrum, while in practice it is treated as a delta function at the effective frequency of each band.}
    \label{fig:spectrum_strom}
\end{figure*}

We compare the reconstructed spectrum of the PWPL approximation with the expected spectrum according to \textsc{cloudy} and the 3 band PWC approximation at column densities $N_\mathrm{HI}=10^{17},~10^{18}~\mathrm{and}~5\times10^{18}\,\mathrm{cm}^{-2}$corresponding to an optical depth at 13.6 eV of $\tau_{13.6}\approx0.64,\,6.4~\mathrm{and}~31.8$, from the source, for the pure H (top) and primoridal H and He mixture (bottom) case in Fig. \ref{fig:spectrum_strom} .
At low column density the medium is fully ionised resulting in an optically thin medium and the reconstructed spectrum agrees well with the expected spectrum.
As radiation propagates further from the source it is absorbed stronger at the low energy end of an ionising band and the flux maximum within one broad band shifts from the low energy side towards the high energy end. 
In the PWPL approximation the fluxes are calculated at the ends of the ionising bands, this allows the PWPL spectral reconstruction to follow the changing slope of the spectrum within the classical ionising bands. 
In case the maximum of the radiation field lies between the two narrow bands the number of photons in the reconstructed spectrum will be lower compared to the expected number in the band, this leads to a an underestimation of the ionisation rate and thus a higher neutral fraction.
This process accumulates with distance from the source and is particularly visible between 24.6 and 54.4 eV (HeI ionising band) in the right panels of the Figure \ref{fig:spectrum_strom}.

In the PWC approximation the spectral shape is encoded in the effective photo-ionisation cross-section, which is representative for an effective photon energy in a given ionisation band. We illustrate this by representing the PWC spectrum (blue lines) in Fig. \ref{fig:spectrum_strom} as scaled down intervals of the original blackbody spectrum for which $\sigma^\mathrm{eff}_{i,j}$ was calculated. In practice, the PWC approximation uses a single intensity value at the effective photon energy. The blackbody pieces in Fig. \ref{fig:spectrum_strom} are scaled such that the integrated flux within one ionising band is conserved.
It becomes obvious that the PWC effective spectral shape does not agree with the expected spectrum once absorption becomes important in shaping the spectrum of the incident radiation field.
The effective PWC spectrum in this test favours photons at the low energy edge of one band, these photons are more likely to ionise the gas as the cross-sections at lower energies are larger. This leads to a greater spatial extend of the ionised bubble compared to the PWPL case, as can be seen in Figs. \ref{fig:Honly} and \ref{fig:HHe}.

We note that the we did not attempt to fully reproduce \textsc{cloudy} results. 
This would require to adjust the cross-sections and reaction rates to those used in \textsc{cloudy}, as attempted in \citet{katz22}. Even then we would potentially be limited by the different approaches to the chemical network, the geometry of the problem (1D in \textsc{cloudy} versus 3D in \textsc{Gasoline2}), the spatial resolution or the spectral range and resolution.
Therefore, we merely use the \textsc{cloudy} calculations as qualitative guide to interpret out test results. 
It is further worth pointing out that there even exist quantitative differences between various \textsc{PDR} codes \citep[e.g.][]{rollig07}.

\subsection{Uniform background in an homogeneous box}\label{sec:BGsources}

This test is designed to show case how we treat background radiation for cosmological zoom-in simulations and the implications for the simulation box size.
We use a low resolution ($32^3$ dark matter and gas particles) cosmological volume of 100 comoving Mpc per side, with $H_0=67.8\rm \,km\,s^{-1}\, Mpc^{-1}$, $\Omega_m=0.3086$, $\Omega_\Lambda=0.6814$ and $\Omega_b=0.048$ cosmology. 
As this test only considers optically thin radiative transfer and no force calculations, the number of particles, their mass and actual position are not relevant, the gas particles merely serve as tracers of the radiation field at their respective location.
In the optically thin case all bands behave qualitatively identical, therefore we choose to only show the results based on the HI-low ($13.6-13.7$ eV) band.

As in the original version of \textsc{Trevr} \citep{grond19}, virtual background sources are distributed along a spiral on a sphere centered on the center of the simulation box with its radius equal to the half of the box width. 
Although the background sources are treated identical to \citet{grond19} it is worth illustrating how the background radiation field is modeled, as a realistic treatment of the UVB has a sever impact on cosmological zoom-in simulations as illustrated in Fig. \ref{fig:MUGS_HIdens} and further discussed in Sect. \ref{sec:mugs}.
In this test we use 128 background sources, which emit light according to the UVB spectrum of \citet{HM12}.
This distribution of background sources is intended to mimic a light emitting shell.
The flux received from the shell at any position within this shell can be calculated by:
\begin{equation}\label{eq:shellflux}
  F(r) = \frac{L}{8\pi\,R}\ln \left(\frac{R+r}{R-r} \right),
\end{equation}
where L is the luminosity of the emitting shell, $R$ is the shell radius and $r$ the distance from the shell's centre.

\begin{figure*}[t]
    \centering
    \includegraphics[trim=1cm 2.5cm 0cm 3cm, width=\columnwidth]{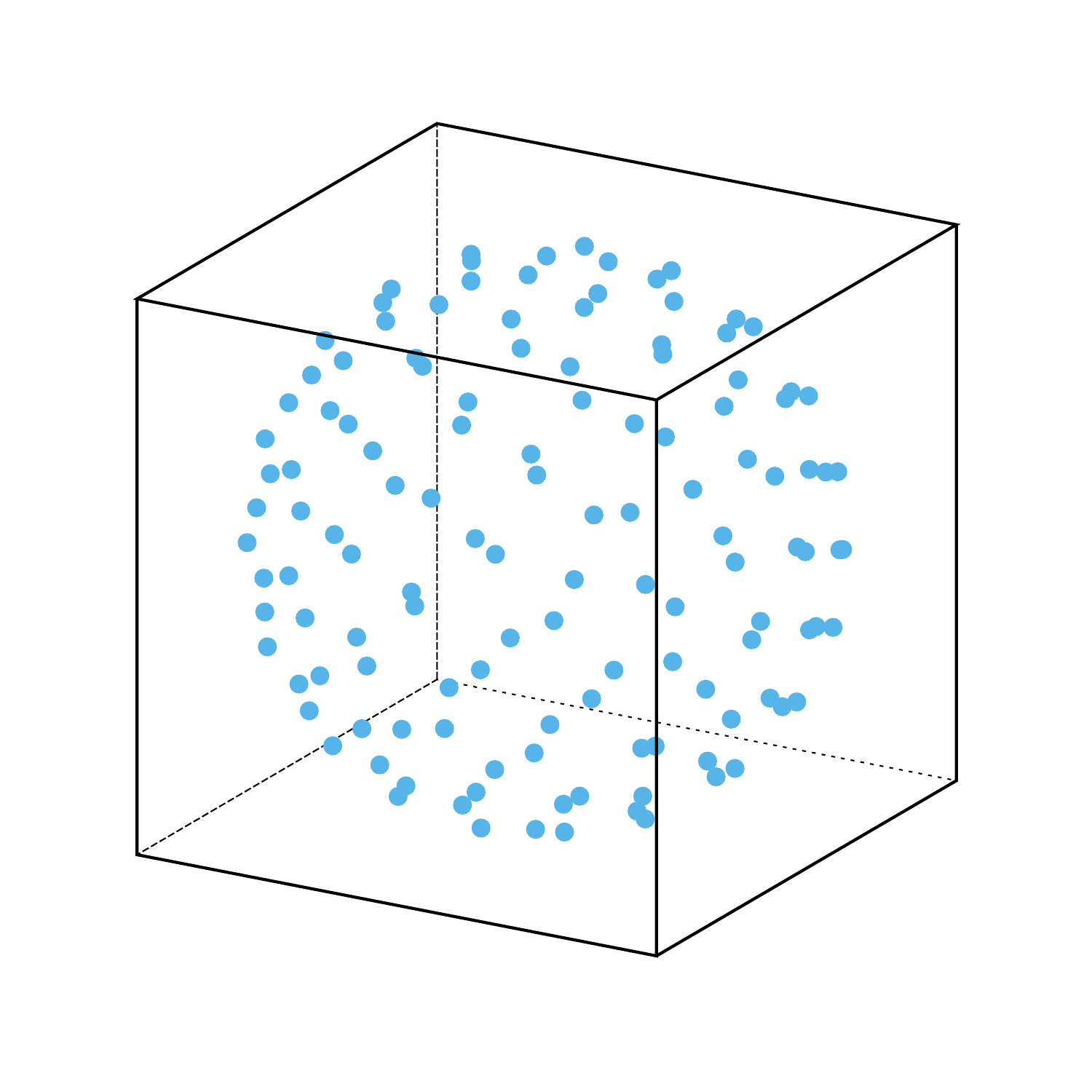}
    \includegraphics[trim=0cm 1cm 0cm 0cm, width=0.9\columnwidth]{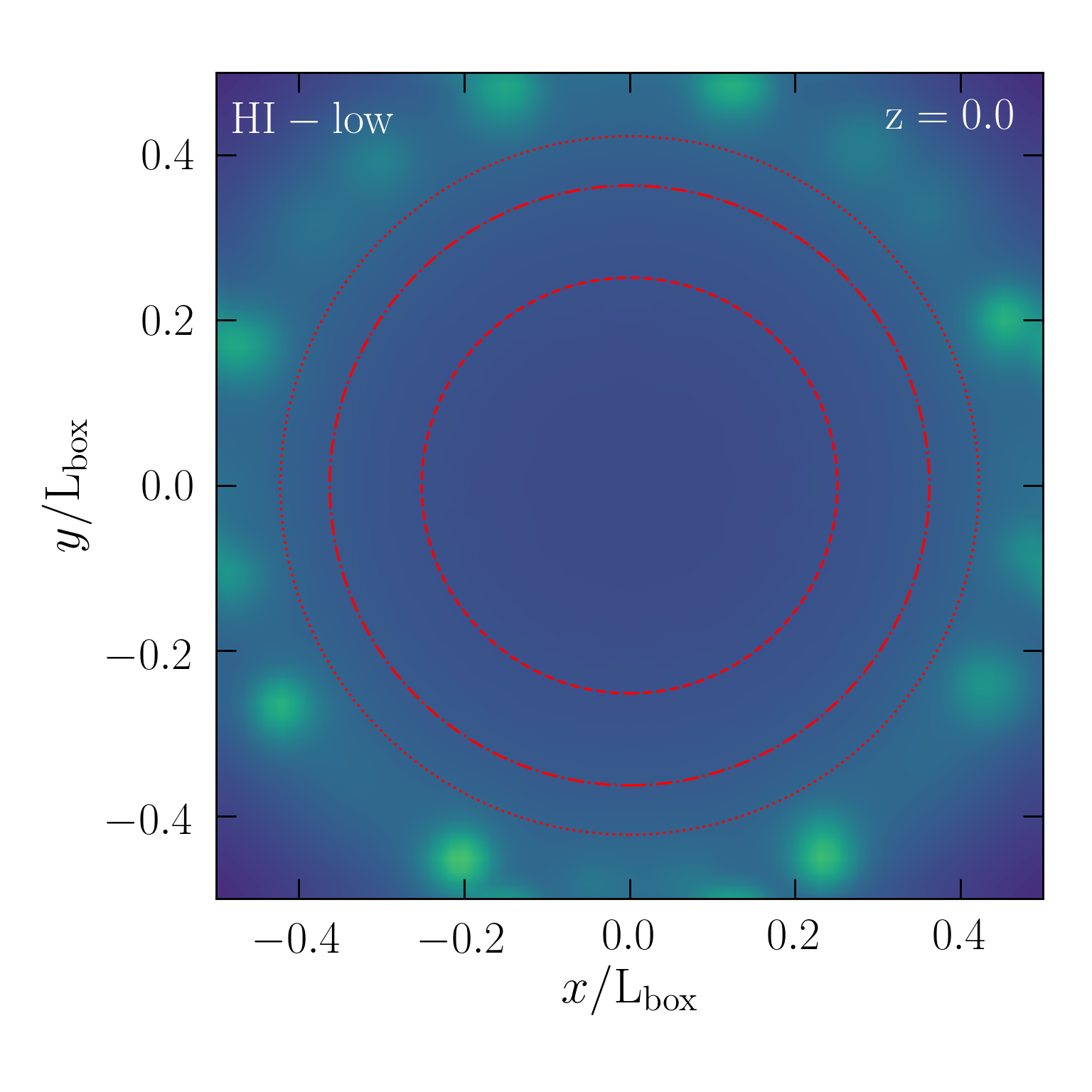}
    \caption{Uniform background test. Left:
    Illustration of the distribution of background sources within the simulation box for 128 background emitters. The virtual background emitters are distributed along a spiral on a sphere centered on the center of the simulation box.
    Right: Incident HI-low ($13.6-13.7$ eV) radiation field as experienced by gas particles in the uniform background test in a thin slice through the center of the simulation box for the \citet{HM12} UVB at redshift 0 and 128 background sources. The dashed, dot-dashed and dotted circles indicate where the radiation field is within 10\%, 25\% and 50\% of the expected central flux, respectively. The corresponding radii are, from inside out, $[0.26, 0.36, 0.43]\times L_\mathrm{box}$.
    }
    \label{fig:bgdist}
\end{figure*}

In the left panel of Figure \ref{fig:bgdist} the distribution of background sources within the simulation volume is illustrated for 128 emitters. 
This behaviour of the radiation field is identical to that seen in \citet{grond19}.
The right panel of Fig. \ref{fig:bgdist} shows the resulting radiation field for our HI-low ($13.6 - 13.7$ eV) band in a thin slice through the centre of the simulation volume. The brighter areas indicate the positions of the background sources. The red circles indicate the regions within which the flux is below 1.5 (dotted), 1.25 (dot-dashed) and 1.1 (dashed) times the central flux, corresponding to $r=[0.43,\, 0.36,\, 0.26]\times L_\mathrm{box}$, respectively. 
In cosmological zoom-in simulations, the relatively small high-resolution region of interest is typically located at the centre of the simulation volume, where the radiation field is only marginally deviating from the expected background intensity and exhibits little scatter.

\section{Post-processing MUGS2 g1536}\label{sec:mugs}

Here we present the results of post-processing seven snapshots of the zoom-in galaxy g1536 form the McMaster Unbiased Galaxy Simulations 2 \citep[MUGS2,][]{keller16}. 

The simulation was originally run with the tree-SPH code \textsc{Gasoline2} \citep{wadsley17} using a WMAP3 $\Lambda$CDM cosmology with $H_0=73\rm \,km\,s^{-1}\, Mpc^{-1}$, $\Omega_m=0.24$, $\Omega_\Lambda=0.76$, $\Omega_b=0.04$ and $\sigma_8=0.76$ \citep{spergel07}.
In the high-resolution zoom region the particle masses are $1.1\times10^6\,\mathrm{M}_\odot$ and $2.2\times10^5\,\mathrm{M}_\odot$ for DM and gas, respectively, and the gravitational softening length was set to 312.5 pc.
These simulations include a non-equilibrium chemical network for H and He, and equilibrium metal line cooling under the influence of a uniform and redshift dependent UVB as described in \citet{shen10}. 
Stars are formed stocastically once a gas particles has become sufficiently cold ($T<1.5\times10^4\,\mathrm{K}$) and dense ($n>9.3\,\mathrm{cm}^{-3}$) at a rate proportional to the local free-fall time:
\begin{equation}\label{eq:sfr}
    \dot{\rho}_\star = c_\star\,\frac{\rho_{gas}}{t_\mathrm{ff}},
\end{equation}
where an efficiency parameter $c_\star=0.1$ is used. 
Supernova (SN) feedback is treated according to the superbubble model \citep{keller14}. In this scheme, gas particles affected by SN feedback temporarily become two phase particles, where one phase represents the SN heated portion of the particle while the other phase is the cold ISM.  The transition between the hot and cold phase is governed by thermal conduction. Both phases have their individual mass, density and specific energy. This allows separate cooling in each phase, eliminating the need for a cooling shut-off as is often used in single phase feedback models to prevent over-cooling.

At $z = 0$ MUGS2 g1536 resembles an almost bulgeless disk galaxy with a total mass of $7\times10^{11}\, \mathrm{M}_\odot$ of which $5.1\times10^{10}\, \mathrm{M}_\odot$ are attributed to gas and $6.0\times10^{10}\, \mathrm{M}_\odot$ to stars, resulting in a baryon fraction $f_{\rm b}=15.9\%$.
This galaxy was used in \citet{keller15} to demonstrate the differences and impact of the superbubble feedback description compared to the blastwave model from \citet{stinson06}, which was used in the original MUGS simulations \citep{stinson10}.

We post-process seven snapshots between $z=6.0\mathrm{~and~}0.0$, with their basic properties listed in Table \ref{tab:mugs}, using \textsc{Gasoline2} and the new ray tracing module \textsc{Trevr2} with our PWPL approximation for the radiation field. 
In order to study the impact of radiation on the gas phase of g1536 we deactivate gravity, hydrodynamics, star formation and stellar feedback in the post-processing runs. Excluding these physical processes disables significant heating processes, such as shock heating and the energy deposition by SN into the ISM.
To counteract the potentially fast cooling of feedback and shock heated gas, we keep the temperature for particles with $T\ge2\times 10^{4}\,\mathrm{K}$ constant but evaluate their ionisation states under the influence of the radiation field.
We evolve the simulation for 1000 steps, equivalent to $\sim13\,\mathrm{Myr}$, in order to reach a steady ionisation state.
We consider stars and a homogeneous UVB as sources of radiation, with their spectra as described in Sect. \ref{sec:instantaneus}. The UVB is represented with 128 virtual sources for which we keep the redshift constant at the value of the respective snapshot. 
The luminosity of each star particle is calculated individually based on the age of stellar population, which we keep fixed throughout the post-processing runs.

\begin{table}
    \centering
    \caption{Properties of the most massive progenitor of MUGS2 g1536 at various redshifts.}
    \begin{tabular}{ccccc}
        \hline\hline
         $z$ & $M_\mathrm{vir}$ & $R_\mathrm{vir}$ & $M_\star$ & $M_\mathrm{gas}$ \\
         & [$\mathrm{M_\odot}$] & [kpc] & [$\mathrm{M_\odot}$] & [$\mathrm{M_\odot}$]  \\
         \hline
        6.0 & 8.3$\,\times\, 10^{9}$  & 84.0 & 1.0$\,\times\, 10^{7}$  & 1.2$\,\times\, 10^{9}$ \\
        3.0 & 1.5$\,\times\, 10^{11}$  & 217.6 & 6.5$\,\times\, 10^{8}$  & 2.1$\,\times\, 10^{10}$ \\
        2.0 & 3.1$\,\times\, 10^{11}$  & 273.3 & 1.6$\,\times\, 10^{9}$  & 4.4$\,\times\, 10^{10}$ \\
        1.0 & 5.2$\,\times\, 10^{11}$  & 299.6 & 4.2$\,\times\, 10^{9}$  & 8.0$\,\times\, 10^{10}$ \\
        0.5 & 6.7$\,\times\, 10^{11}$  & 293.3 & 8.0$\,\times\, 10^{9}$  & 1.0$\,\times\, 10^{11}$ \\
        0.1 & 7.4$\,\times\, 10^{11}$  & 251.2 & 1.5$\,\times\, 10^{10}$  & 1.1$\,\times\, 10^{11}$ \\
        0.0 & 7.5$\,\times\, 10^{11}$  & 236.5  & 1.8$\,\times\, 10^{10}$  & 1.1$\,\times\, 10^{11}$ \\
        \hline
    \end{tabular}
    \label{tab:mugs}
\end{table}

\begin{figure}[t]
    \centering
    \includegraphics[width=\columnwidth]{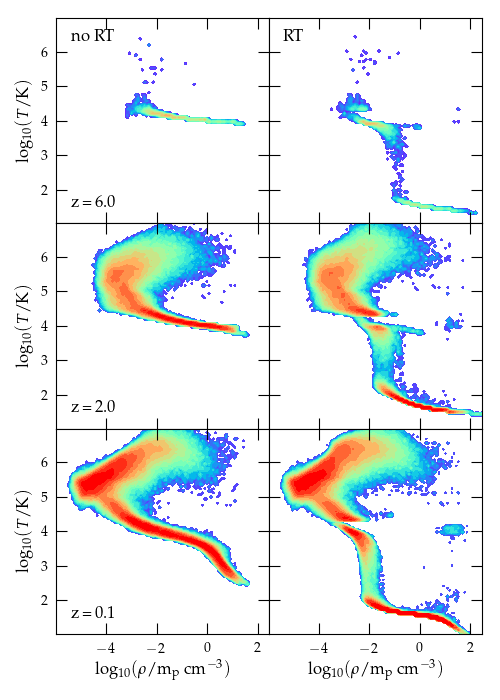}
    \caption{Density-temperature phase-space of the most massive progenitor of g1536 at redshift 6, 2 and 0.1 (from top to bottom). The left panels show the original distribution, i.e. the initial condition for the post-processing run and the right panels show the distribution after the post processing. The horizontal segregation at $T=2\times10^{4}\,\mathrm{K}$, is artificial due to disabling of high temperature cooling (see text for details).
    }
    \label{fig:MUGS_rhoT}
\end{figure}

In Fig. \ref{fig:MUGS_rhoT} we compare the phase-space distribution of gas particles of the most massive progenitor of g1536 before (left) and after post-processing (right) at redshift 6, 2 and 0.1 (from top to bottom). 
At early times ($z=6$) the UVB ionises and heats all of the gas to $T\approx10^4\,\mathrm{K}$ in the initial run, due to its uniform treatment. Simultaneously small burst of star formation create a small amount of SN heated gas with $T>10^4\,\mathrm{K}$. 
During post-processing the radiation emitted by the UVB has to propagate through the IGM and CGM, where large portion of the UVB photons get absorbed before reaching the inner CGM or the ISM. This results in a large portion of the ISM remaining cold ($T<100\,\mathrm{K}$) and neutral, as it is shielded against the UVB. After post-processing the total neutral mass fraction $(M_\mathrm{HI}+M_\mathrm{HeI})/M_\mathrm{gas}$ within the halo of g1536 is 56\% whereas initially it was only 8\%.

\begin{figure}[t]
    \centering
    \includegraphics[width=\columnwidth]{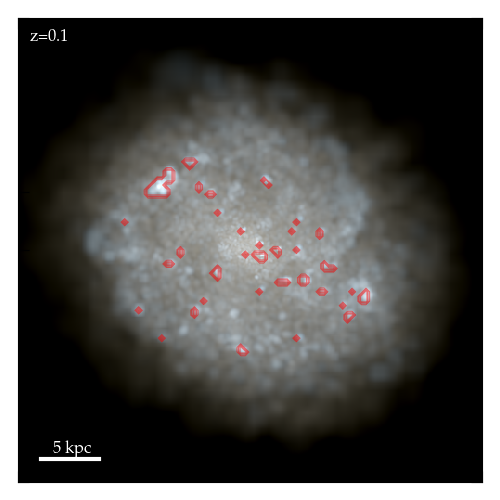}
    \caption{Location of photo-ionised gas in g1536 at $z=0.1$ overlaid onto a mock i, v, b-band composite image of the stellar component, viewed face on. The luminosity of stars are calculated using \citet{margio08} stellar populations, the effect of dust extinction is not included. Particles constituting the HII regions have $\rho\geq1\,\mathrm{cm^{-3}}$ and $7\times 10^3\leq T/\mathrm{K}\leq3\times 10^4$.
    }
    \label{fig:MUGS_starHII}
\end{figure}

At $z=2$, the UVB reaches its maximum strength, keeping the majority of gas ionised in the initial run, while simultaneously the amount of SN heated gas increased. Shortly after this snapshot, g1536 experiences a star burst. First indications of this event can be seen in the formation of a thin extension of the distribution towards high gas density gas. 
Although the UVB is at its strongest, the ISM remains well shielded after post-processing. At temperatures between $7\times10^3$ and $3\times10^4$ K and densities above $1\,\mathrm{cm^{-3}}$ a population of locally ionised HII regions is established around young star clusters.
At $z=0.1$ g1536 is in a phase of ongoing star formation, reflected in the cool ($T<10^4\,\mathrm{K}$) and dense ($n>10\,\mathrm{cm}^{-3}$) gas, which will be converted into stars during the following time steps. In the RT run the ongoing star formation is further reflected in the increased amount of particles in the HII region phase. 
In Fig. \ref{fig:MUGS_starHII} we illustrate the spatial alignment of locally ionised gas with the location of young stars. The red contours show the locations of gas particles with $\rho\geq1\,\mathrm{cm^{-3}}$ and $7\times 10^3\leq T/\mathrm{K}\leq 3\times 10^4$ at $z=0.1$, overlain onto a mock composite image of stars, where brighter blue regions are indicative of young stellar populations.
The horizontal segregation at $T=2\times10^4\,\mathrm{K}$ in the right panels of Fig. \ref{fig:MUGS_rhoT} is an artefact caused by disabling cooling of high temperature gas. 

The persistent existence of cold gas ($T\lesssim100\,\mathrm{K}$) points to a possibly significantly altered star formation history of g1536 in a fully self-consistent cosmological RT simulation. In the original run from \citet{keller15} star formation happens in short ($<7\,\mathrm{Myr}$) bursts with rates up $10\,\mathrm{M_\odot\,yr^{-1}}$ and typically below $5\,\mathrm{M_\odot\,yr^{-1}}$. The phase diagram after post-processing gives rise to the possibility of more continuous SF at higher rates, due to the available amount of cold and dense gas. However, it is difficult to predict how the increased amount of cold gas impact the SFH without self-consistent RT simulations. More cold gas can lead to stronger bursts of star formation, but in turn stellar feedback is also stronger, which could ultimately be more destructive.

\begin{figure}[t]
    \centering
    \includegraphics[width=\columnwidth]{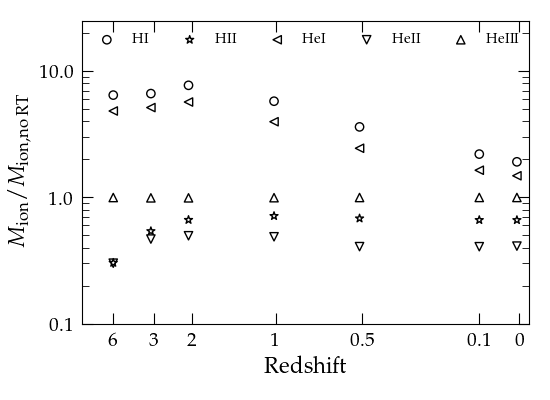}
    \caption{Ion mass relative to the initial ion mass before RT porst-processing, for the considered species of the chemical network, of the most massive progenitor of g1536 at various redshifts.}
    \label{fig:MUGS_mass}
\end{figure}

As already seen in the phase-space diagrams, the amount of cold, and neutral gas, increases significantly after RT post-processing. 
We investigate the change in ion mass for hydrogen and helium relative to the initial ion mass before post-processing in Fig. \ref{fig:MUGS_mass}.
At $z=2$, the increase of neutral hydrogen mass is at its maximum with seven times more neutral hydrogen after post-processing. 
In contrast to the strong change in the mass of neutral ions, the mass of HeIII remains almost unchanged, as the high energy photons required to fully ionise helium have a long mean free path, and are able to penetrate deep into the ISM of g1536.
The ion masses of hydrogen and helium before and after post-processing are provided in Table \ref{tab:mugs1}.
The redshift evolution of the increase/decrease of neutral/ionised mass follows the intensity of UVB, indicating that most of the change is related to the treatment of the UVB. 
In the original MUGS2 run all but the densest and coldest gas in the ISM is ionised by the uniform UVB, whereas in the RT post-processing run gas with $n\gtrsim10^{-2}~\mathrm{cm}^{-3}$ is shielded against the UVB, as can be seen in Fig. \ref{fig:MUGS_rhoT} and was found in previous studies \citep[e.g.][]{gnedin10,rahmati13}.
Stars in the post-processing run ionise parts of the ISM, which have previously been ionised by the uniform UVB, but are not able to entirely compensate for the reduced intensity of the UVB.
The assumption of a spatial uniform and instantaneous UVB without a correction for self-shielding, as used in the original MUGS2 run, underestimates the amount of neutral gas in and around a galaxy.

\begin{table*}
    \centering
    \caption{Neutral and ionised H and He mass of the most massive progenitor of MUGS2 g1536 at various redshifts before and after RT post-processing.}
    \begin{tabular}{c|ccccc|ccccc}
        \hline\hline
         & \multicolumn{5}{l}{before RT \citep{keller16}} & \multicolumn{5}{|l}{after RT post-processing} \\
         $z$ & $M_\mathrm{HI}$ & $M_\mathrm{HII}$ & $M_\mathrm{HeI}$ & $M_\mathrm{HeII}$ & $M_\mathrm{HeIII}$ & $M_\mathrm{HI}$ & $M_\mathrm{HII}$ & $M_\mathrm{HeI}$ & $M_\mathrm{HeII}$ & $M_\mathrm{HeIII}$ \\
         &  [$\mathrm{M_\odot}$] & [$\mathrm{M_\odot}$] & [$\mathrm{M_\odot}$] & [$\mathrm{M_\odot}$] & [$\mathrm{M_\odot}$] & [$\mathrm{M_\odot}$] & [$\mathrm{M_\odot}$] & [$\mathrm{M_\odot}$] & [$\mathrm{M_\odot}$] & [$\mathrm{M_\odot}$] \\
         \hline
        6.0 & 1.0$\,\times\, 10^{8}$  & 8.1$\,\times\, 10^{8}$  & 2.7$\,\times\, 10^{6}$  & 1.5$\,\times\, 10^{7}$  & 2.6$\,\times\, 10^{8}$  & 6.6$\,\times\, 10^{8}$  & 2.4$\,\times\, 10^{8}$  & 1.3$\,\times\, 10^{7}$  & 4.4$\,\times\, 10^{6}$  & 2.6$\,\times\, 10^{8}$ \\
        3.0 & 1.2$\,\times\, 10^{9}$  & 1.5$\,\times\, 10^{10}$  & 3.1$\,\times\, 10^{7}$  & 1.8$\,\times\, 10^{8}$  & 4.8$\,\times\, 10^{9}$  & 8.0$\,\times\, 10^{9}$  & 8.0$\,\times\, 10^{9}$  & 1.6$\,\times\, 10^{8}$  & 8.5$\,\times\, 10^{7}$  & 4.7$\,\times\, 10^{9}$ \\
        2.0 & 1.6$\,\times\, 10^{9}$  & 3.2$\,\times\, 10^{10}$  & 4.4$\,\times\, 10^{7}$  & 2.8$\,\times\, 10^{8}$  & 1.0$\,\times\, 10^{10}$  & 1.2$\,\times\, 10^{10}$  & 2.1$\,\times\, 10^{10}$  & 2.5$\,\times\, 10^{8}$  & 1.4$\,\times\, 10^{8}$  & 1.0$\,\times\, 10^{10}$ \\
        1.0 & 3.4$\,\times\, 10^{9}$  & 5.7$\,\times\, 10^{10}$  & 1.0$\,\times\, 10^{8}$  & 4.5$\,\times\, 10^{8}$  & 1.9$\,\times\, 10^{10}$  & 2.0$\,\times\, 10^{10}$  & 4.1$\,\times\, 10^{10}$  & 4.1$\,\times\, 10^{8}$  & 2.2$\,\times\, 10^{8}$  & 1.9$\,\times\, 10^{10}$ \\
        0.5 & 8.4$\,\times\, 10^{9}$  & 6.9$\,\times\, 10^{10}$  & 2.6$\,\times\, 10^{8}$  & 6.0$\,\times\, 10^{8}$  & 2.4$\,\times\, 10^{10}$  & 3.0$\,\times\, 10^{10}$  & 4.7$\,\times\, 10^{10}$  & 6.3$\,\times\, 10^{8}$  & 2.5$\,\times\, 10^{8}$  & 2.4$\,\times\, 10^{10}$ \\
        0.1 & 1.8$\,\times\, 10^{10}$  & 6.6$\,\times\, 10^{10}$  & 5.4$\,\times\, 10^{8}$  & 5.9$\,\times\, 10^{8}$  & 2.7$\,\times\, 10^{10}$  & 4.1$\,\times\, 10^{10}$  & 4.4$\,\times\, 10^{10}$  & 8.8$\,\times\, 10^{8}$  & 2.4$\,\times\, 10^{8}$  & 2.7$\,\times\, 10^{10}$ \\
        0.0 & 2.3$\,\times\, 10^{10}$  & 6.2$\,\times\, 10^{10}$  & 6.5$\,\times\, 10^{8}$  & 5.4$\,\times\, 10^{8}$  & 2.7$\,\times\, 10^{10}$  & 4.4$\,\times\, 10^{10}$  & 4.1$\,\times\, 10^{10}$  & 9.6$\,\times\, 10^{8}$  & 2.2$\,\times\, 10^{8}$  & 2.7$\,\times\, 10^{10}$ \\
        \hline
    \end{tabular}
    \label{tab:mugs1}
\end{table*}

\begin{figure}[t]
    \centering
    \includegraphics[width=\columnwidth]{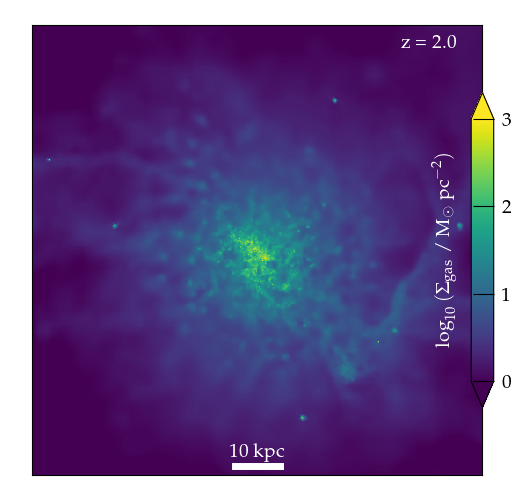}
    \caption{Face-on projected total gas density ($\Sigma_\mathrm{gas}$) in a region with 100 kpc per side centred on the most massive progenitor of g1536 at $z=2$.}
    \label{fig:MUGS_gas}
\end{figure}

Fig. \ref{fig:MUGS_HIdens} shows that most of the additionally formed HI gas is located in the CGM of g1536 at $z=2$. This holds true at all examined redshifts. In the RT post-processing run the UVB light has to propagate inwards from the edge of the simulation box, thereby it is partially absorbed, reducing its ability to penetrate deep into the halo of g1536. 
Further, RT post-processing revealed a large reservoir of neutral gas in a small companion and its associated long (up to $\sim40\,\mathrm{kpc}$) ram pressure stripped tail with $N_\mathrm{HI}>10^{20}~\mathrm{cm}^{-2}$. Although visible in total gas surface density (see Fig. \ref{fig:MUGS_gas}), with $\Sigma_\mathrm{gas}\approx10\,\mathrm{M}_\odot\,\mathrm{pc}^{-2}$, this feature remains hidden in the original MUGS2 run, as the instantaneous and uniform treatment of the UVB ionises and heats all but the densest gas.

We investigate the impact of tree-based RT with cell averaged quantities on the amount of neutral and ionised gas in Appendix \ref{sec:T1compare}.

\section{Discussion}\label{sec:discuss}

In this work we presented a new method to reconstruct the spectral shape of the incident radiation field in RT simulations to accurately calculate photo-ionisation and heating rates. Thereby, we represent the source spectra with two narrow bands at the photo-ionisation edges of H and He and assume that the spectra follow a power-law between the narrow bands.
Having one band just below and one just above the ionisation edge allows to track jumps in intensity across these edges as illustrated in Fig. \ref{fig:spectrum_strom}.
We represent the photo-ionisation cross-sections as piece-wise power-laws, instead of using a single value at an effective frequency. 
These two assumptions enable us to analytically evaluate photo-ionisation and heating rates for any combination of source spectra, and is therefore well suited for hydrodynamic simulations with on-the-fly radiative transfer.
It does not only allow to self-consistently combine radiation from different sources, each with its individual spectrum, but also follows spectral hardening within the typical ionising bands naturally. Furthermore, it is possible to calculate the intensity of the radiation field at any frequency, which is crucial for accurate predictions of the ionisation state of metals.
We explore the impact of the radiation field approximation on the abundance of various metal ions in the CGM in an upcoming work.

The PWPL spectral reconstruction method is applicable in any RT code. We choose to use \textsc{Trevr2} for our tests due to its $N_\mathrm{active}~\mathrm{log}_2(N_\mathrm{src})$ scaling and the low additional cost of performing multi-band RT. Using 10 bands only increased the time required for RT by a factor of $\sim1.3$ compared to a single band.

We validate the PWPL spectral reconstruction method by calculating instantaneous photo-ionisation and heating rates for the \citet{HM12} UVB and an aging SSP.
Thereby, we find excellent agreement between the PWPL method and the full frequency dependent integral over the entire redshift range of the UVB spectrum and considered species, with typical errors bellow 10 per cent. 
For the aging SSP we find typical deviations from the frequency dependent integral of the photo-ionisation rate below 0.1\,dex, within the first $8\,\mathrm{Myr}$ of stellar evolution. 
As the population ages, the error increases to 0.5\,dex over a range of $\sim20\,\mathrm{dex}$ in photo-ionisation rates, at this point the ionisation rates have already decreased by more then 5\,dex from their initil values. Qualitatively the results are similar for the heating rates, where strong deviations arise after about 8, 15, 100 Myr for the HeII, HeI and H heating rates, respectively. 
This is attributed to the discretisation of the stellar spectra into a finite number of bins, regardless of the number of considered bands.
As the population ages the intensity of high energy photons gradually decreases starting at the highest energies (see Fig. \ref{fig:sb99}). 
If the source spectrum only partially reaches into an ionising band, e.g. if the HeII band ($54.4-136$ eV) is truncated at 100 eV, the photons get artificially distributed throughout the whole band. In the PWC approximation the photons would be smeared out over the full energy range of the HeII band, while in the PWPL reconstruction the spectral slope would be calculated between the HeII-low band and the zero-intensity in the HeII-high band.
Therefore, both approximations consider more higher energy photons than are actually present in the input spectrum, resulting in an overestimation of the photo-heating rates.
We have chosen to conserve the number of photons per broad band of the input spectrum, amplifying this effect. If we would have chosen to conserve the total energy per band the errors in the photo-heating rates would be smaller, but at the cost of increasing the errors in the photo-ionisation rates.
However, in applications we do not expect these larger errors in the heating rates to have a significant impact, as the errors are nominally large but at heating rates which are orders of magnitude below the zero-age rates.

We illustrate how the spectrum changes its shape as light travels through the medium of an HII-region, illuminated by a single source (see Fig. \ref{fig:spectrum_strom}).
For optical depths $\tau\lesssim10$ the spectral shape can be recovered well.
As the optical depth rises further the PWPL reconstruction underestimates the number of ionising photons, as the transmitted spectrum is no longer well represented by simple piece-wise power-laws. 
Adding additional narrow bands between the ionisation edges and/or  using a higher order reconstruction are a possible ways to minimise this effect, we leave such improvements for future exploration.
We expect this effect to be less prominent in situations with multiple source with different spectral shapes.

By post-processing MUGS2 g1536, an almost bulgeless, star forming galaxy with $M_\mathrm{vir}=7.5\times10^{11}\,\mathrm{M}_\odot$ at $z=0$, we show the importance of radiative transfer for future (zoom-in) simulations of galaxy evolution.
The inclusion of radiation leads to a realistic multi-phase structure of the ISM, consisting of a cold ($T<100\,\mathrm{K}$), a hot ($T>10^4\,\mathrm{K}$) and a warm ($T\approx10^4\,\mathrm{K}$) phase. The warm phase is co-spatial with recently formed stars which heat and ionise their surrounding gas with their intense UV radiation.
Further, we find large differences in the distribution of neutral gas in the CGM of g1536.
In the comparison run from \citet{keller16}, which utilises spatially uniform photo-ionisation and heating rates for the UVB, only the densest and coldest gas, with recombination rates exceeding the photo-ionisation rates of the UVB, are able to remain neutral. This leads to an almost completely ionised CGM, with only a small amount of neutral gas extending beyond the ISM.
As the radiation emitted by the UVB sources has to propagate through the halo of the galaxy, whereby it is partly absorbed by the intervening matter, a much larger circumgalactic reservoir of neutral gas is present. This neutral gas reservoir extends far beyond the stellar disk, as commonly observed in galaxies \citep[e.g.][]{tumlinson13,wang16} or around $z=2$ quasars \citep[e.g.][]{prochaska13}.
Additionally, we find large neutral fractions in a ram pressure stripped tail of an in-falling dwarf galaxies at $z=2$. The tail is visible in the total gas surface density in the original run (see Fig. \ref{fig:MUGS_gas}), but is completely ionised by the UVB.
Furthermore, we find that a series of small satellite halos can retain neutral gas until redshifts below 2.
These changes in the structure and distribution of neutral gas in the CGM show the importance of radiation and a proper treatment of the UVB for studies of galaxy evolution in general, and in particular for studies of the CGM.

We note that the highlighted differences before and after RT post-processing are only to be considered illustrative for the potential changes in galaxy simulations if RT is included and the UVB is treated appropriately. All the presented changes in the structure of the galaxy and its CGM indicate a significantly altered star formation history.
In preliminary tests we saw star formation rates increased by a factor of 5 to 10 compared to the original run without RT.
This in turn effects the stellar feedback and radiation field, and thus the dynamics of the whole system.

These changes and their impact on the gas properties of MUGS2 g1536 show the need for fully self-consistent cosmological (zoom-in) RT simulations in order to further develop our understanding of galaxy evolution. Thereby the RT code and the used approximations made for the radiation field and related quantities must allow for the combination of sources with different spectra and include a realistic treatment of UVB - as it is possible with our new PWPL spectral reconstruction method.

In upcoming works we will present such cosmological zoom-in simulations of galaxy evolution with on-the-fly RT, including a realistic treatment of the UVB, employing our PWPL spectral reconstruction method.

\begin{acknowledgements}
  BB and SS acknowledge support from the Research Council of Norway through NFR Young Research Talents Grant 276043. SS also acknowledge support from the European High Performance Computing Joint Undertaking (EuroHPC JU) and the Research Council of Norway through the funding of the SPACE Centre of Excellence (grant agreement N0 101093441). JW is supported by a Discovery Grant from NSERC of Canada. Parts of the computations were performed on resources provided by Sigma2 - the National Infrastructure for High Performance Computing and Data Storage in Norway and on the Niagara supercomputer at the SciNet HPC Consortium. SciNet is funded by: the Canada Foundation for Innovation; the Government of Ontario; Ontario Research Fund - Research Excellence; and the University of Toronto.
\end{acknowledgements}

\bibliographystyle{aa}
\bibliography{10BandRT}

\begin{appendix}
\section{The impact of cell averaged tree-based RT}\label{sec:T1compare}

\citet{wadsley23} showed, in idealised tests, how linearly averaging absorption in tree cells can lead to significant errors, especially when these are vastly different, as is typical for a clumpy medium. A small amount of highly opaque material can lead to an unrepresentative high opacity in a tree cell.

In order to asses the impact of tree-based RT with cell averaged transmission in an astrophysically interesting case we performed two additional post-processing runs of MUGS2 g1536 at $z=2.0$ using the original version of \textsc{Trevr} \citep{grond19}. 
The first run \textsc{Trevr} \textit{noRef} used default settings and no refinement. The second run \textsc{Trevr} \textit{ref1} enforces refinement of tree cells once their optical depth exceeds a set threshold of $\tau_\mathrm{ref}=1.0$. We refer the reader to \citet{grond19} for details of the refinement process in \textsc{Trevr}.
We consider the results of the \textsc{Trevr} \textit{ref1} as the most accurate and thus as reference run for this comparison.

To quantify the impact of the different methods we choose to compare the mass of ionised H and He after 300 RT post-processing steps, due to the high computational costs of \textsc{Trevr} with refinement. At this stage the neutral and ionised H and He masses have converged to within 10\% of the final equilibrium value.
Table \ref{tab:testBox} and \ref{tab:testHalo} list the neutral and ionised H and He mass and their fraction within the whole simulation box and the most massive progenitor halo of MUGS2 g1536, respectively.

\begin{figure}
    \centering
    \includegraphics[width=\columnwidth]{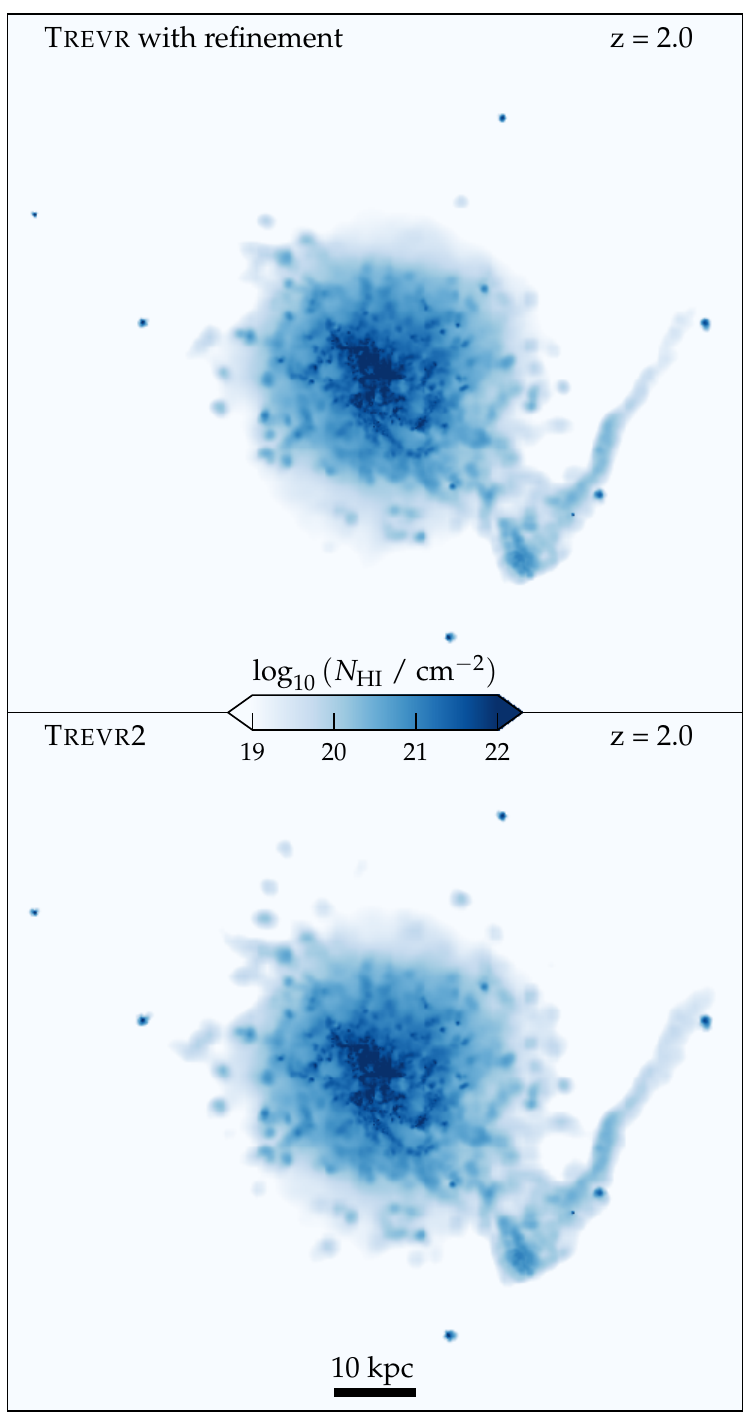}
    \caption{Face-on projected hydrogen density ($N_\mathrm{HI}$) of the reference run \textsc{Trevr} \textit{ref1} (top) and \textsc{Trevr2} with default settings (bottom) after 300 RT steps.}
    \label{fig:refTest}
\end{figure}

Overall the three runs are in excellent agreement with each other. On the simulation box scale the total mass of the neutral and ionised species only differ by a few percent between the three runs, considering the mass fraction the maximal difference is only 0.3\%.
Within the most massive progenitor halo the neutral and ionised fractions of H are within $\approx1\%$ of the reference run (\textsc{Trevr} \textit{ref1}), while the He fractions are with $\approx0.1\%$ of each other.
Despite on a low level, there is a trend observable in these data, \textsc{Trevr} without refinement (\textit{noRef}) produces the least amount of neutral gas and \textsc{Trevr2} has consistently the highest neutral fraction.
We illustrate the similarities between the reference run \textsc{Trevr} \textit{ref1} and \textsc{Trevr2} in Fig. \ref{fig:refTest}, where we show neutral hydrogen column density ($N_\mathrm{HI}$).

This behaviour shows how well suited \textsc{Trevr2} with transmission averaging is for zoom-in simulations of galaxies, it produces near identical ionised and neutral masses while being up to 100 times faster than \textsc{Trevr} with refinement.

\begin{table*}
    \centering
    \caption{Neutral and ionised mass and fraction of H and He within the simulation box of MUGS2 g1536 at redshift 2.0.}
    \begin{tabular}{l|cccccccccc}
         \hline\hline
         Post-processing & $M_\mathrm{HI}$ & $M_\mathrm{HII}$ & $f_\mathrm{HI}$ & $f_\mathrm{HII}$ & $M_\mathrm{HeI}$ & $M_\mathrm{HeII}$ & $M_\mathrm{HeIII}$ & $f_\mathrm{HeI}$ &$f_\mathrm{HeII}$ & $f_\mathrm{HeIII}$ \\
         run & [$M_{\odot}$] & [$M_{\odot}$] & [\%] & [\%] & [$M_{\odot}$] & [$M_{\odot}$] & [$M_{\odot}$] & [\%] & [\%] & [\%] \\
         \hline
         \textsc{Trevr} \textit{noRef}  & 1.251$\times10^{10}$ & 1861$\times10^{11}$ & 6.30 & 93.70 & 2.439$\times10^{8}$ & 2.101$\times10^{8}$ & 1.494$\times10^{10}$ & 1.58 & 1.37 & 97.05 \\
         \textsc{Trevr} \textit{ref1}   & 1.272$\times10^{10}$ & 1.859$\times10^{11}$ & 6.41 & 93.59 & 2.497$\times10^{8}$ & 2.079$\times10^{8}$ & 1.493$\times10^{10}$ & 1.62 & 1.35 & 97.03 \\
         \textsc{Trevr2} & 1.334$\times10^{10}$ & 1.853$\times10^{11}$ & 6.72 & 93.28 & 2.620$\times10^{8}$ & 1.913$\times10^{8}$ & 1.494$\times10^{10}$ & 1.70 & 1.24 & 97.05 \\
         \hline
    \end{tabular}
    \label{tab:testBox}
\end{table*}

\begin{table*}
    \centering
    \caption{Neutral and ionised mass and fraction of H and He of the most massive progenitor of MUGS2 g1536 at redshift 2.0.}
    \begin{tabular}{l|cccccccccc}
         \hline\hline
         Post-processing & $M_\mathrm{HI}$ & $M_\mathrm{HII}$ & $f_\mathrm{HI}$ & $f_\mathrm{HII}$ & $M_\mathrm{HeI}$ & $M_\mathrm{HeII}$ & $M_\mathrm{HeIII}$ & $f_\mathrm{HeI}$ &$f_\mathrm{HeII}$ & $f_\mathrm{HeIII}$ \\
         run & [$M_{\odot}$] & [$M_{\odot}$] & [\%] & [\%] & [$M_{\odot}$] & [$M_{\odot}$] & [$M_{\odot}$] & [\%] & [\%] & [\%] \\
         \hline
         \textsc{Trevr} \textit{noRef}  & 1.072$\times10^{10}$ & 2.279$\times10^{10}$ & 31.99 & 68.01 & 2.103$\times10^{8}$ & 1.600$\times10^{8}$ & 1.053$\times10^{10}$ & 2.00 & 1.52 & 96.48 \\
         \textsc{Trevr} \textit{ref1}   & 1.100$\times10^{10}$ & 2.251$\times10^{10}$ & 32.83 & 67.17 & 2.168$\times10^{8}$ & 1.582$\times10^{8}$ & 1.015$\times10^{10}$ & 2.06 & 1.50 & 96.44 \\
         \textsc{Trevr2} & 1.137$\times10^{10}$ & 2.215$\times10^{10}$ & 33.93 & 66.07 & 2.242$\times10^{8}$ & 1.483$\times10^{8}$ & 1.016$\times10^{10}$ & 2.13 & 1.41 & 96.46 \\
         \hline
    \end{tabular}
    \label{tab:testHalo}
\end{table*}
\end{appendix}

\end{document}